\documentstyle[aps,twocolumn]{revtex}
\begin{document}
\title{Spectral properties of the 2D Holstein polaron}
\author{H. Fehske$^1$, J. Loos$^2$  and G. Wellein$^1$ } 
\address{
$^1$Physikalisches Institut, Universit\"at Bayreuth,
D--95440 Bayreuth, Germany\\
$^2$Institute of Physics, Czech Academy of Sciences, 16200
Prague, Czech Republic}
\date{Bayreuth, 30 June 1997}
\maketitle
\input{epsf}
\def\gsim{\hbox{$\lower1pt\hbox{$>$}\above-1pt\raise1pt\hbox{$\sim$}$}}
\def\lsim{\hbox{$\lower1pt\hbox{$<$}\above-1pt\raise1pt\hbox{$\sim$}$}}
\def\ep{\varepsilon_p}
\def\o{\omega}
\def\ho{\omega_{\!o}}
\def\gs{\tilde{g}}
\def\ta{t^{\ast}}
\def\bd{{\vec{d}}}
\def\bm{{\vec{m}}}
\def\bms{{\vec{m}^{\prime}}}
\def\bmss{{\vec{m}^{\prime\prime}}}
\def\bl{{\vec{l}}}
\def\bh{{\vec{h}}}
\def\bk{{\vec{K}}}
\def\bks{{\vec{K}^{\prime}}}
\def\mP{{\mit \Phi}}
\def\mS{{\mit \Sigma}}
\def\cA{{\cal{A}}}
\def\cC{{\cal{C}}}
\def\cD{{\cal{D}}}
\def\cF{{\cal{F}}}
\def\cG{{\cal{G}}}
\def\cH{{\cal{H}}}
\def\cK{{\cal{K}}}
\def\cN{{\cal{N}}}
\def\cS{{\cal{S}}}
\def\cT{{\cal{T}}}
\def\cU{{\cal{U}}}
\def\cW{{\cal{W}}}
\def\cZ{{\cal{Z}}}
\begin{abstract}
The two--dimensional Holstein model is studied by means of direct Lanczos
diagonalization preserving the full dynamics and quantum nature of phonons.
We present numerical exact results for the single--particle spectral function,
the polaronic quasiparticle weight, and the optical conductivity.
The polaron band dispersion is derived both from exact diagonalization
of small lattices and analytic calculation of the polaron self--energy. 
\end{abstract}
\section{Introduction}
Research on polarons has recently gained renewed interest on account 
of the observation of polaronic effects in several important classes 
of materials with perovskite related structure, 
such as the bismuthates ($\rm Ba_{1-x}K_xBiO_3$), the  
high--$T_c$ cuprates (e.g., $\rm La_{2-x}Sr_xCuO_4$), 
the non--metallic nickelates  ($\rm La_{2-x}Sr_xNiO_4$),
or the colossal magnetoresistive manganites 
(e.g., $\rm La_{1-x}Ca_xMnO_3$)~\cite{SAL95}. 
Despite a more than a half--century of theoretical and experimental
study, our understanding of the formation and transport properties 
of polarons and bipolarons is still incomplete.
Even the very fundamental problem of a single tight--binding 
electron coupled locally to a set of non--interacting Einstein oscillators
(Holstein model~\cite{Ho59a}) 
has not been solved exactly, i.e., no satisfactory 
description of the full spectral properties has been obtained so far.      
That is because the standard analytical techniques, based, e.g., 
on variational approaches~\cite{Em73} or on  
weak--coupling~\cite{Mi58} and strong--coupling 
adiabatic~\cite{Ho59a} and non--adiabatic~\cite{LF62,Ma95} 
perturbation expansions, fail to tackle this complicated many--body
problem precisely in the physically most important transition region, 
where the highly non--linear ``self--trapping'' process     
from an essentially delocalized quasi--free charge carrier 
at weak electron--phonon (EP) interaction to a ``quasi--localized''  
small polaron at strong coupling takes place. Note that 
the same problem arises in strongly coupled exciton--phonon 
systems as well~\cite{LR74}.

As an attempt to close the gap between the
weak and strong EP coupling limits, in this paper 
we investigate the Holstein model on finite lattices mainly  
employing approximation--free diagonalization techniques
which, at the moment, provide the only reliable tool for
studying polarons close to the crossover regime. 

Explicitly we write the  Holstein Hamiltonian 
\begin{eqnarray}
\cH&=&-t\sum_{\langle ij\rangle} ( c_i^{\dagger} c_j^{} 
+ c_j^{\dagger} c_i^{})
+\ho\sum_i  ( b_i^{\dagger} b_i^{}+ 
\mbox{\small $\frac{1}{2}$})\nonumber\\ 
&&-\sqrt{\ep\ho}\sum_i
( b_i^{\dagger}  + b_i^{})  c_i^{\dagger} c_i^{}\,,
\label{ED1}
\end{eqnarray}
where $ c_i^{[\dagger]}$ and $ b_i^{[\dagger]}$ are the annihilation 
[creation] operators of a (spinless) fermion and a boson (phonon) 
at Wannier site $i$, respectively. In (1), the free electron transfer ($t$)
is restricted to nearest--neighbour (NN) pairs ($\langle ij \rangle$),
the phonons are treated within harmonic approximation, and the EP 
term introduces a coupling  between the local electron density   
and the dispersionsless optical phonon mode. Denoting by
$g$ the dimensionless EP interaction constant, $\ep=g^2\ho$ is the 
Lang--Firsov  strong--coupling polaron binding energy 
with $\ho$ as the bare phonon frequency. 

The single--electron Holstein model has been studied extensively 
as a paradigmatic model for polaron formation in systems with
dominant short--range electron--lattice interactions. 
The principal result is that irrespective of the adiabaticity ratio, 
$\alpha=\ho/t$, small polarons will be formed in a $D$--dimensional system 
provided that the {\it two} conditions 
$\lambda=\ep/2Dt>1$ and $g^2>1$ are fulfilled (see Fig.~1).
In the weak--coupling (adiabatic) limit, 
one expects (from scaling arguments)
a quasi--free electron behaviour for $D>1$,
while in a 1D system the carrier state becomes polaronic 
at arbitrary small $\lambda$ (``large'' polaron). 
Previous exact diagonalization (ED) work has concentrated on the 1D 
case~\cite{ZS92,RT92,Ma93,KR94,AKR94,FRWM95,WRF96,BGPV96,CSG97,Ro97,MR97}, 
where, however, the calculations were limited to either  
very small clusters, rather weak EP coupling 
$(\lambda < 1)$  or to the adiabatic limit  $(\ho=0)$. 
On the other hand, the most interesting effects will  
be expected if the characteristic electronic and phononic 
energy scales are not well separated  ($\lambda\sim 1;\;\alpha\sim 1$). 
For example, very recent numerical investigations, performed 
in the crossover region for 1D chains with up to 20 sites,
indicate that the polaronic band structure markedly deviates
from a simple tight--binding dispersion~\cite{St96,WF97,PB97}.  

Our purpose here is to extend these studies to 2D systems and
explore the spectral properties of the 2D Holstein model, whereby 
the focus is on the {\it intermediate} EP coupling and phonon frequency
regime. Section 2 presents the ED results 
for the single--particle spectral function and 
the optical conductivity. 
In Sect. 3 we outline an analytical approach 
based on an ``incomplete'' Lang--Firsov transformation,  
to calculate second--order corrections to the polaronic self--energy.
Section 4 contains a discussion of the polaronic band dispersion,
where the ED data are used  to test the validity of the proposed theory.
The principal results are summarized in Sec.~5.

\section{Numerical approach}
Performing direct diagonalizations of the Holstein model on  finite
($N$--site) square lattices with periodic boundary conditions (PBC), 
we exploit the translational invariance by the use 
of symmetrized basis sets in the tensorial product Hilbert space
of electronic and phononic  states 
\begin{equation}
\{|\vec{K};s \rangle\}=\sum_{i=1}^N
\frac{\mbox{e}^{\,i\vec{K}\vec{R}_i}}{\sqrt{N}} 
{\cal T}_{\vec{R}_i} (|1,0,\ldots,0\rangle_{el}
\otimes \{|s\rangle_{ph}\})\,.
\label{ED2}
\end{equation}  
Here ${\cal T}_{\vec{R}_i}$ describes the allowed lattice translations and
$\vec{K}$ is the {\it total} momentum of the coupled EP system.
Since the Hilbert space associated with the phonons is infinite
even for a finite system, we apply a truncation 
procedure~\cite{II90,WRF96} restricting ourselves to phononic states
$|s\rangle_{ph}=\prod_{i=1}^N
\frac{1}{\sqrt{n_i^s!}}\left(b_i^\dagger\right)^{n_{i}^{s}}\,|0\rangle_{ph}$
with at most $M$ phonons. That means, we  take into account  all 
$m$--phonon states, where $m=\sum_{i=1}^N n_i^s \le M$,  $n_i^s\in [0,m]$ 
and $1\le s \le D_{ph}^{(M)}$. Here $ D_{ph}^{(M)} =(M+N)!/M!N!\,$
denotes the dimension of the phononic Hilbert space. 
Then a $\bk$--symmetrized state of the Holstein model is 
given as
\begin{equation}
|{\mit \Psi}_{\bk}^{}\rangle = \sum_{m=0}^M
\sum_{\bar{s}=1}^{\bar{S}(m)} c_{\bk}^{m,\bar{s}} \,
|\vec{K};m,\bar{s} \rangle\,,
\label{ED2a}
\end{equation} 
where $\bar{S}(m)=(N-1+m)!/(N-1)!m!$. 

Needless to say, 
that we have carefully to check for the convergence of both the
ground--state energy,$E_0(M)$, and the phonon distribution,
$|c^m|^2(M)=\sum_{\bar{s}}  |c_{\bk=0}^{m,\bar{s}}|^2$, 
as a function of the maximal number of phonons 
retained~\cite{WRF96}. Implementing an improved Lanczos algorithm on 
parallel computers, we are able to calculate the ground--state 
properties of systems with a total dimension of about $10^7$.
To obtain information about dynamical properties, we combine the
Lanczos diagonalization with the Chebyshev recursion and maximum entropy
methods, which are very efficient for high--energy resolution 
applications~\cite{Siea96}. 
\subsection{Single--particle spectral function}
By examining the dynamical properties of polarons, of particular 
importance is whether a quasiparticle--like excitation exists 
in the spectrum. In order to discuss this issue we have 
evaluated the wave--vector resolved spectral density function  
\begin{equation}
\cA_{\bk}(E) = \sum_n |\langle {\mit\Psi}_{n,\bk} 
\,|\,c_{\bk}^{\dagger}\,|\,0\rangle|^2\,\delta ( E-E_{n,\bk})\,,
\label{ED3}
\end{equation}
where $\,c_{\bk}^{\dagger}$ creates an electron with momentum
$\bk$; $|0\rangle$ is the vacuum state (containing no phonons). 
For a finite system, the eigenvalues $E_{n,\bk}$ and 
eigenstates $|{\mit\Psi}_{n,\bk} \rangle$
in the single--particle subspace can be obtained 
from ED.  As stated above, we prefer to use a moment
expansion method in order to calculate $\cA_{\bk}(E)$  directly.

In Fig.~2, $\cA_{\bk}(E)$ is displayed for two different
EP couplings $\lambda=0.125$ (a--c) and  $\lambda=1.25$ (d--f)  
(corresponding to the weak and intermediate--to--strong coupling 
situations, respectively)  at the allowed $\bk$--vectors of the 
18--site square lattice.  
To visualize the intensities (spectral weights) connected with
the various peaks (excitations) we also have shown the 
integrated density of states
\begin{equation}
\cN_\bk(E)=\int_{-\infty}^{E} dE^{\prime}\cA_{\bk}(E^{\prime})\,. 
\label{ED4}
\end{equation}
Let us first consider the weak--coupling case.  
Owing to the ``hybridization'' of electron and phonon degrees of freedom,
we found in each $\bk$--sector excitations being separated from the 
ground state by an energy of about $\ho$. 
As can be seen from $\cN_\bk(E)$,  however, 
these predominantly phononic states have only a small admixture of
electronic character, i.e., the main spectral weight 
is still located at energies that correspond 
to the bare tight--binding levels $\epsilon_{\bk}=
-2t (\cos K_x + \cos K_y)$. Other (satellite) peaks appear due 
to the existence of higher excited multi--phonon states.   
Increasing the EP interaction, we notice significant changes in
the spectral function as $\ep$ exceeds half the bare bandwidth.
Most notably, a band of polaronic states with low spectral
weight  evolves. This so--called small polaron band becomes 
extremely narrow and is well--separated from the rest of the 
spectra in the strong--coupling limit ($\lambda\gg 1$).
At the same time spectral weight is redistributed to the 
high--energy part of the spectra. 

To elucidate the nature of the ground state in more detail, in Fig.~3~a 
we have plotted the phonon distribution in the ground state 
at several EP interaction strengths and phonon frequencies. 
The results unambiguously confirm the importance of multi--phonon
states especially in the strong--coupling
regime $(g^2>>1)$, where $|c^m|^2$ becomes the usual Poisson distribution
for small polarons. Here the phonons will heavily dress the electron
and as a result the bare electron no longer behaves like a 
well--defined quasiparticle. In fact, this is exactly what 
has been observed in Fig.~2~d, where the residue of the quasiparticle 
peak at $\bk =0$,
\begin{equation}
\cZ^{(c)} = |\langle {\mit  \Psi}_{0,\bk}^{} | c_{\bk}^{\dagger} 
| 0 \rangle |^2_{\bk=0}\,,
\label{ED5}
\end{equation} 
is weakened. Then the question arises whether
one can construct an appropriate quasiparticle operator
having large spectral weight at the lowest pole in the spectrum. 
This point was intensively discussed in the context of 
strongly correlated electron models (e.g., for the t--J model) 
for the case of an additional hole injected in the antiferromagnetic 
ground state of the undoped system~\cite{Da94} and, more recently,  
similar ideas have been addressed for strongly coupled EP systems 
as well~\cite{Ro97,SPS97}. Of course, in the parameter regime
$\lambda\gg 1$, $\alpha > 1$  the small polaron defined through 
the Lang--Firsov transformation~\cite{LF62} (cf. Sec.~3) is a 
good quasiparticle. This was explicitly demonstrated in
recent ED work~\cite{Ro97}. 

To tackle the intermediate coupling and frequency regime, we construct  
a composite polaron operator with momentum $\bk$ on the basis of the 
phonon distribution function,
\begin{equation}
d^{\dagger}_{\bk}= \sqrt{\frac{1}{N}} \sum_{i=1}^{N}
\mbox{e}^{i\bk \vec{R}_i} c_i^{\dagger}\sum_{m=0}^{M}
\sqrt{\frac{|c^m(M)|^2}{m!}}\;\left(b^{\dagger}_i\right)^m\,,
\label{ED6}
\end{equation}
keeping in mind that the polaronic state is characterized by 
strong on--site EP correlations.  
In Fig.~3~b, we have compared the quasiparticle weight factor $Z^{(d)}$,
computed by the use of the dressed electron operator~(\ref{ED6}), with 
the spectral weight simply given by~(\ref{ED5}). As expected, 
the results for $Z^{(d)}$ are actually very close to those 
for $Z^{(c)}$ at weak enough EP coupling. Clearly, in this case 
the phonon distribution is sharply peaked at the 
zero--phonon state (cf. Fig.~3~a) and the particle 
behaves as a nearly free electron. By contrast, if  one increases 
the EP coupling strength, both electrons and phonons loose their 
own identity by forming polarons. 
Concomitantly we observe less ``electronic character'' 
of the polaronic quasiparticle and, indeed, it can be shown 
analytically that (in the bulk limit) $Z^{(c)}$ vanishes exponentially at 
very large couplings $(\lambda \gg 1)$.  
On the other hand, although care must be taken 
in the interpretation of our finite--cluster data 
for $\cZ$, the large value found for  $Z^{(d)}$ suggests
that the proposed  $d$--operator correctly describes the 
phonon dressing of the quasiparticle even in the transition regime.
Accordingly the position of the maxima in the distribution function 
provides an estimate of the (most probable) number of phonons contained 
in the ``phonon-cloud'' of the small polaron. 
\subsection{Optical conductivity}
In linear response theory the real part of the optical 
conductivity is given by~\cite{Da94} 
\begin{equation}
\mbox{Re}\sigma_{xx}(\o ) = \cD\delta(\o) +\sigma^{reg}_{xx}(\o)\,,
\label{ED7}
\end{equation}
where $\cD$ denotes the Drude weight and the second term 
(sometimes called incoherent or regular part of the conductivity) 
can be written in a spectral representation as
\begin{equation}
\sigma^{reg}_{xx}=\frac{e^2\pi}{N}\sum_{m\neq 0}
\frac{|\langle {\mit \Psi}_0^{} | \hat{j}_x^{(p)} |  {\mit \Psi}_m^{}
       \rangle |^2}{E_m-E_0} \;\delta(\o-E_m+E_0)\,.
\label{ED8}
\end{equation}
For the Holstein model the (paramagnetic) current density
operator has the form  
\begin{equation}
\label{ED9}
\hat{j}_x^{(p)} =it\sum_{i}( c_{i}^{\dagger}
 c_{i+x}^{} - c_{i+x}^{\dagger}c_{i}^{})\,.
\end{equation}
According to the  $f$--sum rule the integrated conductivity  
is related to the ground--state expectation value of the kinetic
energy operator
\begin{equation}
\cH_t =-t\sum_{\langle ij \rangle }( c_{i}^{\dagger}
 c_{j}^{} + c_{j}^{\dagger}c_{i}^{})\,.
\label{ED10}
\end{equation}
By introducing the $\o$--integrated weight $\cS^{reg}=\cS^{reg}(\infty)$, 
\begin{equation}
\cS^{reg}(\o)=\int_0^{\o} d\o^{\prime} 
\,\sigma_{xx}^{reg}(\o^{\prime})\,,
\label{ED11}
\end{equation}
and integrating in $\o$ both terms on the r.h.s. of~(\ref{ED7}), 
one arrives at 
\begin{equation}
 \cS^{tot}=\frac{\pi e^2 }{4}\langle -\cH_t \rangle = \frac{\cD}{2} + 
\cS^{reg}\,.
\label{ED12}
\end{equation}

As pointed out by Emin~\cite{Emi93}, there are two simple  
limits in which absorption associated with photoionization 
of Holstein polarons is well understood and the optical conductivity
can be calculated analytically. The first one is 
the weak--coupling case, where the absorption coefficient falls
monotonically with increasing applied frequency. 
In the opposite strong--coupling (small polaron) limit, 
the optical properties are dominated by (incoherent) small polaron
hopping processes accompanied by multi--phonon absorptions and
emissions (i.e., by non--diagonal transitions~\cite{Mah90}). As a result 
the absorption is peaked about $\o \sim 2\ep$, whereby, in contrast
to the case of large (Fr\"ohlich--type) polarons, the low--frequency
side is evaluated above the high--frequency side of the absorption 
peak~\cite{Emi93}. Recently, for the weak--, intermediate-- and 
strong--coupling regimes absorption spectra are obtained with 
numerical calculation of the finite--frequency part of the 
optical conductivity of (1D)  
finite--size Holstein models with up to 
four--sites~\cite{AKR94b,CSG97}. 

Since it is not expected that the dimensionality (cluster size) 
plays a crucial role in the extreme small polaron limit,
in the discussion of the optical response we again focus on 
the crossover region. Results are presented in Fig.~4. 
The upper panel shows the frequency dependence of the optical conductivity 
(omitting the Drude contribution, i.e., the intra-band transitions). 
The most interesting qualitative feature of the absorption spectrum 
seems to be the strongly asymmetric lineshape. 
The low--frequency peak structure may be easily understood in 
connection with the single--particle spectra. 
For example, an inspection of Fig.~2~d--f verifies that the first/second 
group of peaks correspond to inter-band transitions from the
$\bk=(0,0)$ ground state to states with finite 
momenta, triggered by one/two phonon absorption
processes ($\o\sim E_{0,\bk}-E_0 + n\ho$, where 
$\ho=1.5$ and  $(E_{0,\bk}-E_0)\sim 0.5$
for $\bk$=(1,1), (2,0), (2,2), (3,1), (3,3) [see also Fig.~5~a below]).  
In principle such processes are possible even at arbitrarily small 
EP coupling because there is always a finite overlap 
of the corresponding wave--functions.  
However, the peak strength rapidly decreases 
if the EP coupling becomes weaker. This can be seen 
from the inset of Fig.~4~a, by comparing the integrated weight 
$\cS^{reg}(\o)$ in the dissipative part of $\sigma (\o)$.

To analyze the role of the EP coupling strength on the optical
response in more detail, we have displayed $\cS^{reg}$ together with 
the total sum rule (kinetic energy) in  Fig.~4~b.  
$\cS^{tot}$ is directly related to the effective polaronic hopping 
amplitude, $t_{p,eff}=\cS^{tot}/\pi e^2$,
which can be taken as a measure of a polaron's mobility~\cite{WRF96,FRWM95}. 
Containing both coherent and incoherent transport processes, 
$t_{p,eff}$ substantially differs from the (exponential)
polaron band renormalization factors obtained analytically in
the adiabatic Holstein and non--adiabatic Lang--Firsov cases~\cite{AKR94}
(cf. Appendix). Fig.~4~b clearly shows the crossover
from a nearly free electron, characterized by a $t_{p,eff}$ that is 
only weakly reduced from its non--interacting value  
($t_{p,eff}(\lambda =0)=1$), to a less mobile small 
(Holstein/Lang--Firsov) polaron in the 
(adiabatic/non--adiabatic) strong--coupling limit.
Concomitantly, in the weak--coupling regimes  
nearly all the spectral weight stays in the Drude  part.  
As previously found for the (1D) four--site 
Holstein cluster~\cite{CSG97}, the sharp decrease 
of  $\cS^{tot}$ in the crossover region is driven by
the fall of the Drude weight. 
By contrast the optical absorption due to
inelastic scattering processes,  described by the regular part of 
the optical conductivity, becomes enhanced in the transition region.
Of course,  $\cS^{reg}$  also decreases as $\lambda$ increases in the 
extreme strong--coupling regime.

\section{Theoretical approach}
In this section we treat the intermediate--to--strong EP coupling 
regime analytically, applying the Green's function formalism 
advocated by Schnakenberg~\cite{Sc66}
to the polaron problem. 
We start with the Holstein Hamiltonian~(\ref{ED1})
transformed by the {\it incomplete} Lang--Firsov transformation $\cU=
\exp\{\gamma g\sum_i(b_i^{\dagger}-b_i^{})\,c_i^{\dagger}c_i^{}\}$, 
$\tilde{\cH}=\cU^{\dagger}\cH\cU$ \cite{LF62,Feea94}:
\begin{equation}
\tilde{\cH}=\eta\sum_i c_i^{\dagger}c_i^{} -\sum_{i,j} 
\cC_{ij} c_i^{\dagger}c_j^{} +\ho\sum_i  ( b_i^{\dagger} b_i^{}+ 
\mbox{\small $\frac{1}{2}$})\,,
\label{L1}
\end{equation}
where
\begin{eqnarray}\label{L2}
\eta&=&-\ep\gamma(2-\gamma)-\mu\,,\\\label{L3}
\cC_{ii}&=& g\ho(1-\gamma) ( b_i^{\dagger} +b_i^{})\,,\\\label{L4}
\cC_{ij}&=&t\mP_{\langle ij\rangle}=
t\exp\{-\gamma g(b_i^{\dagger} - b_i^{} - b_j^{\dagger}
+ b_j^{} )\}\,.
\end{eqnarray}
In (\ref{L1}), the (variational) parameter $\gamma$ measures the degree
of the dynamical polaron effect $(0\leq\gamma\leq 1)$; 
$\mu$ is the chemical potential. 

The Green's function equations of motion deduced from $\tilde{\cH}$
lead to a set of coupled equations for  generalized 
polaron  ($c$--$c$)  and ``mixed''  ($c$--$\cC$)
 Green's functions; the equation for the generalized polaron
Green's function may be converted into an equation 
for the generalized self--energy and solved by iteration 
(see Refs.~\cite{Sc66,Lo94,KB62} for details).
In the second step of iteration, the polaron self energy is obtained as
\begin{eqnarray}\label{L5}
\lefteqn{\mS^{(2)}(\bm_1 \tau_1;\bm_2\tau_2)=
-\langle\cC_{\bm_1\bm_2}\rangle \delta(\tau_1-\tau_2)}\\
&&+\sum_{\bms \bmss}\cG(\bms\tau_1;\bmss\tau_2)
[\langle\cT_{\tau}\cC_{\bm_1\bms}(\tau_1)\cC_{\bmss\bm_2}(\tau_2)\rangle
\nonumber\\
&& \hspace*{4cm}-\langle\cC_{\bm_1\bms}\rangle\langle\cC_{\bmss\bm_2}\rangle]
\,,\nonumber
\end{eqnarray}
where the $\tau$--dependence of the boson operators $b^{[\dagger]}_i$
and the statistical averages are to be determined using the Hamiltonian 
of independent  local oscillators (the last term on the r.h.s. of (\ref{L1})).
Introducing the abbreviations $\cW(\bm_1 \bms; \bm_2 \bmss; \tau)$
for the expression in the square brackets of (\ref{L5}) [the $\bm_i$, 
$\bms$, and  $\bmss$ refer to lattice vectors, and $\tau=\tau_1-\tau_2$], 
the Fourier transformation of~(\ref{L5}) gives the self--energy 
equation in the space of Brillouin zone $\bk$--vectors 
and Matsubara frequencies $i\o_\nu =i(2\nu+1)\pi/\beta$, i.e.,  
\begin{eqnarray}\label{L6}
\lefteqn{
\mS^{(2)}_{\bk}(i\o_\nu)=-\sum_{\bm_2-\bm_1}\langle\cC_{\bm_1\bm_2}\rangle
\mbox{e}^{i\bk(\bm_2-\bm_1)}}\\
&&+\sum_{\bm_2-\bm_1}\mbox{e}^{i\bk(\bm_2-\bm_1)}\frac{1}{N}
\sum_{\bks,\varsigma;\bms\bmss}\mbox{e}^{i\bks(\bms-\bmss)}
\cG_{\bks}(i\o_\varsigma) 
\nonumber\\
&&\times\frac{1}{\beta}\int_0^\beta d\tau\mbox{e}^{i(\o_\nu-\o_\varsigma)\tau}
\cW(\bm_1\bms;\bm_2\bmss;\tau)\,.\nonumber
\end{eqnarray}
Owing to the properties of $\cC_{ij}$, the first approximation 
to the self energy, $\mS^{(1)}_\bk$, 
given by the first term on the r.h.s. of~(\ref{L6}), 
is non--zero only for elementary translations $\bh=\bm_2-\bm_1$
connecting the site $\bm_1$ with the NN sites $\bm_2$; consequently,
the $\bk$--dependence of this term is given by $2(\cos K_x+\cos K_y)$.
The terms obtained in the second step of iteration yield, besides
for the case   $\bm_2-\bm_1=0$, non--zero contributions only for 
$\bh^{\prime}=\bm_2-\bm_1$ (vector from $\bm_1$ to next NN site $\bm_2$)
and  $\bm_2-\bm_1=2\bh$ leading to the $\bk$--functions $4\cos K_x \cos K_y$
and  $2(\cos 2K_x+\cos 2K_y)$, respectively.
In fact, denoting the NN sites to  $\bm_1$ and  $\bm_2$ by  
$\bm_1 + \bd_1$ and  $\bm_2 +\bd_2$, respectively, the non--vanishing
$\cW(\bm_1\bms;\bm_2\bmss;\tau)$ are given by
\begin{equation}
\cW(\bm,\bm+\bd_1;\bm,\bm+\bd_1;\tau)=t^2\cF_1(\tau;2\gs^2)
\label{L7}
\end{equation}
for $\bm_1=\bm_2=\bm$ and $\bd_1=\bd_2$,
\begin{equation}
\cW(\bm,\bm+\bd_1;\bm,\bm+\bd_2;\tau)=\ta t\cF_1(\tau;\gs^2)
\label{L8}
\end{equation}
for $\bm_1=\bm_2=\bm$ and $\bd_1\neq\bd_2$,
\begin{equation}
\cW(\bm_1,\bm_1+\bd_1;\bm_2,\bm_2+\bd_2;\tau)=\ta t\cF_1(\tau;\gs^2)
\label{L9}
\end{equation}
for $\bm_2-\bm_1=\bh^{\prime},\,2\bh$ and simultaneously
$\bm_1+\bd_1=\bm_2+\bd_2$,
\begin{eqnarray}
\cW(\bm,\bm+\bd_1;\bm,\bm;\tau)&=&\cW(\bm,\bm;\bm,\bm+\bd_2;\tau)\nonumber\\
&&\hspace*{-2cm}=\ta \gamma(1-\gamma)\ep\cF_2(\tau)
\label{L10}
\end{eqnarray}
for $\bm_2-\bm_1=0$,
\begin{eqnarray}
\cW(\bm_1,\bm_2;\bm_2,\bm_2;\tau)&=&\cW(\bm_1,\bm_1;\bm_2,\bm_1;\tau)
\nonumber\\ 
&&\hspace*{-2cm}=-\ta \gamma(1-\gamma)\ep\cF_2(\tau)
\label{L11}
\end{eqnarray}
for $\bm_2-\bm_1=\bh$, and 
\begin{equation}
\cW(\bm,\bm;\bm,\bm;\tau)=\ep (1-\gamma)^2\ho\cF_3(\tau)\,.
\label{L12}
\end{equation}
Here, $\ta=t\exp\{-\gs^2\coth \vartheta\}$,  $\gs = \gamma g$
and the functions $\cF_{1,2,3}$ are defined as follows
\begin{eqnarray}
\lefteqn{
F_1(\tau,\kappa)=\exp\{-\kappa\coth\vartheta\}}\\ &&\times
\Big[2\sum_{s=1}^{\infty}
I_s(\zeta)\cosh[s(\vartheta-\ho\tau)]+I_0(\zeta)-1\Big]\nonumber\,,
\label{L13}
\end{eqnarray}
\begin{eqnarray}
\label{L14}
F_2(\tau)&=&\sinh(\vartheta-\ho\tau)/\sinh\vartheta\,,\\
F_3(\tau)&=&\bar{n}\exp\{\ho\tau\}+(\bar{n}+1)\exp\{-\ho\tau\}\,,
\label{L15}
\end{eqnarray}
where $\vartheta=\beta\ho/2$, $\bar{n}=[\exp \{\beta\ho\}-1]^{-1}$,
$\zeta=\kappa/\sinh \vartheta$, and $I_s(\zeta)$ denote the modified
Bessel functions.

Deriving the above expressions of $\cW$, 
the averages $\langle\ldots\rangle$ were
evaluated applying~\cite{Lo85}
\begin{equation}
\left\langle \mbox{e}^{Xb} \mbox{e}^{-Yb^\dagger}\right\rangle
= \exp\bigg\{\frac{-XY}{1-\exp\{-\beta\ho\}}\bigg\}\,. 
\label{16}
\end{equation} 
In particular, the functions originating 
from the $\cT_\tau$--ordered products of $\cC_{ii}$ and 
$\mP_{ij}$ were calculated as $\epsilon$--derivatives of 
the generating functional $\langle \cT_\tau (\exp\{\epsilon[
b_i^\dagger (\tau_1)+ b_i^{} (\tau_1)]\}\mP_{ij}(\tau_2))\rangle$ 
(at $\epsilon =0$).    
After inserting~(\ref{L7})--(\ref{L12}) into~(\ref{L6}), the 
$\tau$--integration is easily accomplished. As a next step, the summation
over the Matsubara frequencies $i\o_\varsigma$ has to be carried out.
For this aim, the Green's function in~(\ref{L6}) will be expressed by means of the spectral function $\cA_{\bks}(\o^\prime)$, i.e.,   
\begin{equation}
\label{L17}
\cG_{\bks}(i\o_\varsigma)=\int_{-\infty}^{\infty}\frac{d\o^\prime}{2\pi}
\cA_{\bks}(\o^\prime)\frac{1}{i\o_\varsigma-\o^\prime}\,,
\end{equation}
and $\cA_{\bks}(\o^\prime)=2\pi\delta(\o^\prime - \xi_\bks)$ will be 
assumed, where $ \xi_\bks=\eta+\mS^{(1)}_\bks$ is given by the first 
approximation of the quasiparticle energy. 
Further approximations that will be made
in the explicit calculations are based (i) on the assumption
$\vartheta \gg 1$ (low--temperature approximation) and 
(ii) on the limitation to negligible carrier concentration. 
In this way we obtain 
\begin{eqnarray}
\label{L18}
\lefteqn{
\frac{1}{N}\sum_{\bks,\varsigma}\cG_{\bks}(i\o_\varsigma)
\frac{1}{\beta}\int_0^\beta d\tau\mbox{e}^{i(\o_\nu-\o_\varsigma)\tau}
F_1(\tau,\kappa)}\\
&&=\sum_{s=1}^\infty\mbox{e}^{-\kappa}\frac{\kappa^s}{s!}
\frac{1}{N}\sum_{\bks}\left[
\frac{n_B(s\ho)+n_F(\xi_{\bks})}{i\o_\nu-\xi_{\bks}+s\ho}\right.\nonumber\\
&&\hspace*{3cm}\left.+\frac{n_B(s\ho)+1-
n_F(\xi_{\bks})}{i\o_\nu-\xi_{\bks}-s\ho}\right]\,.\nonumber
\end{eqnarray}
In view of (i) and (ii), we can neglect both  $n_B(s\ho)$ and $n_F(\xi_\bks )$ 
in~(\ref{L18}). The contributions to the self energy  
determined by the functions 
$\cF_2(\tau)$ and  $\cF_3(\tau)$ are treated in a quite analogous
manner. Collecting all non--zero contributions to the r.h.s of~(\ref{L6})
and performing the analytical continuation $i\o_\varsigma\to\bar{\o}
=\o + i\delta$ ($\delta\to 0^+$), the following low--temperature
formula results:  
\begin{eqnarray}
\label{L19}
\lefteqn{
\mS^{(2)}_{\bk}(\bar{\o})=-2\ta(\cos K_x +cos K_y)}\\
&&+zt^2\sum_{s=1}^{\infty}\mbox{e}^{-2\gs^2}\frac{(2\gs^2)^s}{s!}
\frac{1}{N}\sum_{\bks}\cG^{(1)}_{\bks}(\bar{\o}-s\ho)\nonumber\\
&&+\ta t\sum_{s=1}^{\infty}\mbox{e}^{-\gs^2}\frac{\gs^{2s}}{s!}
\frac{1}{N}\sum_{\bks,\bd_1\neq\bd_2}\!\!\!
\mbox{e}^{i\bks(\bd_1-\bd_2)}\cG^{(1)}_{\bks}(\bar{\o}-s\ho)
\nonumber\\
&&+2\ta t[4\cos K_x \cos K_y +\cos 2 K_x +\cos 2 K_y]\nonumber\\
&&\;\;\;\;\times \sum_{s=1}^{\infty}\mbox{e}^{-\gs^2}\frac{\gs^{2s}}{s!}
\frac{1}{N}\sum_{\bks}\cG^{(1)}_{\bks}(\bar{\o}-s\ho)\nonumber \\
&&+2\ta\gamma(1-\gamma)\ep \frac{1}{N}
\sum_{\bks,\bh}\mbox{e}^{i\bks \bh} \cG^{(1)}_{\bks}(\bar{\o}-\ho)
  \nonumber\\
&&-4 \ta \gamma(1-\gamma)\ep (\cos K_x +\cos K_y) 
 \frac{1}{N}
\sum_{\bks}\cG^{(1)}_{\bks}(\bar{\o}-\ho) 
 \nonumber\\
&&+ (1-\gamma)^2\ep\ho\frac{1}{N}\sum_{\bks}
\cG^{(1)}_{\bks}(\bar{\o}-\ho)\,.\nonumber
\end{eqnarray}
$\cG^{(1)}_{\bk}(w)=[w-\xi_\bk]^{-1}$ denotes the first--order  
(Hartree--Fock) Green's function for independent quasiparticles
having band energies $\xi_\bk$; $z$ is the NN coordination number. 
  
The polaron self energy given by (\ref{L19}) will be used to determine the
polaron band dispersion. 
Namely, the quasiparticle energy is obtained as a function of $\bk$
by solving
\begin{equation}
\label{L20}
\o=\eta+ \mbox{Re} \mS_{\bk}^{(2)}(\bar{\o})\,.
\end{equation}

In the strong--coupling case $\gamma$ near~1 and an extreme polaron
band narrowing are expected; the corresponding (crude) approximation
consists in the omission of  terms arising owing to the ``residual'' 
EP interaction given by $\cC_{ii}$ and the neglect of $(\o-\xi_\bks )$
with respect to the energy scale of the Poisson distribution of the oscillator
energy fluctuations $s\ho$. Then the polaron band dispersion  
is explicitly obtained from~(\ref{L19}) as follows        
\begin{eqnarray}
\label{L21}
\lefteqn{E_{\bk}(\gamma)/4t=-\lambda\gamma (2-\gamma) -\frac{1}{\alpha} 
\Big\langle\frac{1}{s}\Big\rangle_{\kappa=2\gamma^2 g^2}^{}}
\\[0.1cm]&& -\frac{\mbox{e}^{-\gamma^2 g^2}}{2} (\cos K_x +\cos K_y)
- \frac{\mbox{e}^{-\gamma^2 g^2}}{2\alpha}\Big\langle\frac{1}{s}
\Big\rangle_{\kappa=\gamma^2 g^2}^{}\nonumber\\&&
\;\;\;\;\times[4\cos K_x \cos K_y +\cos 2 K_x +\cos 2 K_y]\,,\nonumber
\end{eqnarray}
where $\langle 1/s \rangle_\kappa$ means the average of 
$s^{-1}$ with respect to the Poisson distribution with parameter $\kappa$.
In the limit $\gamma\to 1$, (24) becomes equivalent to the result of 
second--order Rayleigh--Schr\"odinger perturbation theory. 

In the general case, the self--energy is a functional of the function
${\mit \Omega}_{\bks}=\o-\xi_{\bks} =\o -\eta -\mS^{(1)}_{\bks}$, i.e.,
on the basis of~(\ref{L19}), (\ref{L20}) we have
\begin{equation}
\label{L22}
{\mit \Omega}_{\bk}= \mbox{Re} \mS^{(2)}_{\bk}[{\mit \Omega}_{\bks}]
-\mS^{(1)}_{\bk}\,,
\end{equation}
and the dispersion relation becomes
\begin{equation}
\label{L23}
E_{\bk}={\mit \Omega}_{\bk} -\ep\gamma(2-\gamma)  + \mS^{(1)}_{\bk}\,.
\end{equation}
\section{Polaron band structure} 
We are now in the position to discuss the single--particle 
band dispersion of the 2D Holstein model. In the first place, 
we can extract the $\bk$--dependence of the so--called ``coherent'' 
energy band, $E_{\bk}$, from the position of the lowest peak 
in each spectral function $\cA_{\bk}$ using the ED data of Sec.~2.1
(cf. Fig.~2~d--f). This has be done in our previous work~\cite{WF97}. 
In the weak--coupling case and for phonon frequencies less than the 
bare electronic bandwidth, we have found a nearly unaffected tight--binding
band near the band center and a practically flat region 
at larger momenta. As a result the coherent bandwidth $\Delta E=
\sup_{\bk} E_{\bk} -  \inf_{\bk} E_{\bk}$ is roughly given by $\ho$.

The results for the intermediate--to--strong EP coupling regime 
are shown in Fig~5. Besides the well--known narrowing of the 
bandwidth ($\Delta E/8t = 0.0814$), there are at least 
two features worth mentioning in the crossover region (see Fig.~5~a).   
Firstly, the exact band structure differs significantly from a 
simple tight--binding band having the same bandwidth. 
Moreover, because of further than NN hopping processes 
generated by the EP interaction, the enhancement of the 
effective mass is weakened near $\bk =0$.
Secondly, it should be noted that the ``flattening'' of 
the band structure, discussed above for weak EP coupling,  
survives to relatively large EP interactions   
(even though $\ho > 2\Delta E$). At this point we 
would like to stress that the finite--size effects are rather small, 
i.e. although the data points belong to different system sizes
we found a remarkably smooth behaviour of $ E_{\bk}$.  

In order to check for the applicability of the theory outlined in
Sec.~3, we have compared $E_{\bk}$ 
given by~(\ref{L21}) with the exact results, 
where the parameter $\gamma$ was determined to reproduce the correct 
ground--state energy. As can be seen from Fig.~5~b, we found a
surprisingly good agreement of the theoretical curve  
with the numerical data.
It is natural to ask why  our strong--coupling approach is so good  
even in the intermediate EP coupling situation. 
First of all,  
the second--order corrections to the self--energy involve  
longer--ranged polaron hopping processes due to
(multi--) phonon absorption and emission. As mentioned above 
such processes are of particular importance in the crossover
region. Thus we expect an even better agreement 
including higher--order corrections~\cite{PB97}. 
The more important point, however, seems to be the 
use of the incomplete Lang--Firsov
transformation in the derivation of~(\ref{L21}).
Obviously, both the Lang--Firsov formula 
($E_{\bk}=-\ep-2t\exp\{-g^2\}(\cos K_x + \cos K_y)$) and standard 
second--order strong--coupling perturbation theory (SCPT)
fail to describe the band dispersion  correctly. 
Moreover, as might be expected,  both approximation 
considerably overestimate the renormalization of coherent bandwidth 
for intermediate EP couplings and phonon frequencies.
A discussion of the strong--coupling case will be given separately in the 
Appendix. 
\section{Conclusions} 
To summarize, in this paper we have performed a highly complete numerical  
analysis of the 2D Holstein model studying the effect of 
polaron formation. The emphasis was on the intermediate coupling regime,
i.e., the transition region from nearly free electrons to small polarons.  
The calculations are carried out by exact 
diagonalization of finite systems including the full dynamics 
of phonons. Results are unbiased and allow to 
test a new theoretical approach applied to 
the calculation of the polaron self--energy   
as well as previously proposed theories. 
Our main results are the following:
\begin{itemize}
\item[(i)] In the 2D Holstein model a continuous transition from a
nearly free electron to a small polaron takes place if both 
criteria, $\lambda > 1$ and $g^2>1$, are satisfied, i.e., $\lambda$ and
$g^2$ are not by themselves independent parameters~\cite{WF97,CSG97}.
\item[(ii)] The small polaron state is basically a multi--phonon state 
characterized by strong on--site electron--phonon correlations
making the particle susceptible to self--trapping~\cite{WRF96}.
\item[(iii)] Depending on the adiabatic ratio $\alpha$, there are two 
types of small polaron states in the very strong--coupling regime:
the adiabatic Holstein polaron and the 
non--adiabatic Lang--Firsov polaron
(cf. Appendix). By the use of the correct polaron theory, i.e. the adiabatic 
Holstein approximation and the canonical Lang--Firsov approach with
appropriate corrections, one obtains an excellent estimate of the
coherent bandwidth in both adiabatic and non--adiabatic 
regimes~\cite{KR94,AM94}
\item[(iv)] The kinetic energy loss as a function of the EP 
coupling strength indicates that the ``width'' of the crossover region
is rather small (broad) in the adiabatic (non--adiabatic) 
regime~\cite{FRWM95,CSG97}. 
\item[(v)] The single--particle spectral function shows the evolution
of a well-separated narrow small polaron band with low spectral weight.
On the other hand, composite particle operators, properly dressed by
dynamical phonons, give large spectral weight in the spectral function, 
i.e., they are much better fundamental excitations of the 
systems~\cite{Ro97,SPS97}.
\item[(vi)] At intermediate EP coupling strengths and phonon frequencies
the effective polaronic band dispersion deviates substantially from a simple 
tight--binding cosine band due to further than nearest--neighbour
ranged hopping processes generated by the EP interaction. The 
proposed theoretical strong--coupling approach, based on an incomplete
Lang--Firsov transformation, yields a satisfactory description of the 
polaron band structure, whereas the standard Lang--Firsov and 
strong--coupling perturbation theories fail to give the correct energy
dispersion (coherent bandwidth). 
\item[(vi)] The lineshape of the optical absorption spectra is 
highly asymmetric in the weak--to--intermediate coupling regime,
whereby the optical response on the low--energy side of the absorption
peak exceeds that on the high--energy side~\cite{Emi93,AKR94}.\\
\end{itemize}
{\it Acknowledgements.} 
This work was supported in part by the Deutsche 
Forschungsgemeinschaft through SFB~279 B5, and by the Grant Agency of
Czech Republic, Project No. 202/96/0864. Numerical calculations 
were carried out at the LRZ M\"unchen, the HLRZ J\"ulich,  
and the HLR Stuttgart. 
\section*{Appendix} 
Here we would like to add some comments on the nature of the
small polaron state in the strong--coupling limit of the Holstein model.
There is some confusion in the literature, particularly about the
use of the correct theory of polarons (with appropriate corrections, 
cf. Refs.~\cite{Ma93,KR94,AKR94}). 
This may be originated by the fact, that in the extreme strong--coupling 
limit both the adiabatic Holstein~\cite{Ho59a} and non--adiabatic 
Lang--Firsov~\cite{LF62} formulae, obtained from expansions in powers of
$\alpha\ll 1$~\cite{Ho59a,Ra85} and $1/\lambda \ll 1$~\cite{Go82},
respectively, yield the same exponential band renormalization: 
$\exp\{-g^2\}$. However, the pre--exponential factors are different for any
value of $\lambda$~\cite{AKR94}, and the generalized Holstein formula 
derived by Alexandrov and Mott~\cite{AM94} also contains essential 
corrections of the exponent, $g^2\to \bar{g}^2$, 
at intermediate--to--strong EP couplings. 

These findings are supported by exact diagonalizations, 
which we have performed for the 1D Holstein model at fixed $g^2$
but different adiabatic ratios $\alpha$. The comparison of the 
exact results with the theoretical approaches is presented in Fig.~6.
In the strong--coupling anti--adiabatic limit 
$(g^2=5,\;\alpha =10)$, where our theory coincides with the SCPT, 
we obtained an excellent agreement with the exact dispersion
for all $\bk$.
Moreover, the simple Lang--Firsov formula works perfectly well
in the determination of the coherent bandwidth $\Delta E_{LF}\simeq
\Delta E= 0.027$. This is in contrast to the adiabatic  
strong--coupling case $(\alpha=0.8,\;\lambda=2,\;g^2=5)$.
Here the non--adiabatic Lang--Firsov and SCPT expressions 
are found to be quite inadequate, e.g., they underestimate
the exact bandwidth  $\Delta E= 0.0562$ by more than a factor of two.
Although the theory of Sec.~3 improves this result to some extent
$(\Delta E(\gamma =0.97)= 0.0362)$, a much better estimate 
of the bandwidth is achieved using the generalized 
Holstein formula~\cite{AM94}: $\Delta E_{Ho} =0.0577$.
In any case, in real solids the coherence of these small--polaron band states
will be rapidly destroyed by thermal fluctuations and/or disorder effects.
\bibliography{ref}
\bibliographystyle{phys}
\newpage
\figure{FIG. 1. Phase diagram of the 2D Holstein model obtained for a 256--site
square lattice using the  variational Lanczos approach proposed by the authors 
in Ref.~\protect\cite{FRWM95}. This technique, based on an inhomogeneous
variational Lang--Firsov transformation, correctly reproduces the 
adiabatic $(\alpha\ll 1$) and anti--adiabatic  $(\alpha\gg 1$), 
weak--  $(\lambda \ll 1$) and strong--coupling  $(\lambda,\,g^2 \gg 1$) 
limits. We stress that there is no true phase transition between the nearly
free electron and small polaron states, 
i.e., the transition line only indicates 
the  crossover region.}
\figure{FIG. 2. Single--particle spectral function $\cA_{\bk}(E)$ (thin lines) 
and partial integrated density  of states $\cN_{\bk}(E)$  (bold lines)
for the 2D  Holstein model with 18 sites at $\lambda=0.125$, $\alpha =0.8$ 
(left column, {\bf a}--{\bf c}) and 
$\lambda =1.25$, $\alpha =1.5$ (right column, {\bf d}--{\bf f}).
The $\bk$--values are given in units of $\pi/3$.  
Here and in the following all energies are measured in units of $t$.}
\figure{FIG. 3. Weight of the $m$--phonon states $|c^m|^2$ ({\bf a}) 
and wave--function renormalization $\cZ^{(c)}$ [$\cZ^{(d)}]$ 
of a single electron ({\bf b}) in the $\bk =0$
ground state of the 2D Holstein model with ten sites. The residue
of the ``quasiparticle'' pole shown in {\bf b} was calculated
using both the bare (open symbols) and dressed electron (filled symbols) 
operators. For further explanation see text.}
\figure{FIG. 4. Optical absorption in the 2D Holstein model. Using PBC,  
results for the regular part of the conductivity, $\sigma_{xx}^{reg}$ and 
$\cS^{reg}(\o)$, are obtained for the 18--site lattice with $M=10$ phonons
({\bf a}), whereas the various contributions to the sum rule are calculated 
on a ten--site lattice with  $M=16$ phonons ({\bf b}).} 
\figure{FIG. 5. Band dispersion of the 2D Holstein model along the highly
symmetric directions of the Brillouin zone. 
In {\bf a}, the exact results obtained for finite 
lattices with $N=16$ and 18 sites are compared with a rescaled   
tight--binding band (dotted line).
The solid line is a least--squares fit to the ED data. 
In {\bf b}, the chain--dashed line gives the band structure 
according to the theory developed in Sec.~3. Long-- and short--dashed
lines are the results of the Lang--Firsov approximation and the second--order
SCPT, respectively.}
\figure{FIG. 6. Small polaron band dispersion for the 1D Holstein model.
Exact results are compared with the predictions of different 
analytical approaches in the adiabatic ($\alpha=0.8$) and anti--adiabatic
($\alpha=10$) strong--coupling cases ($g^2=5,\;\lambda>1$).}
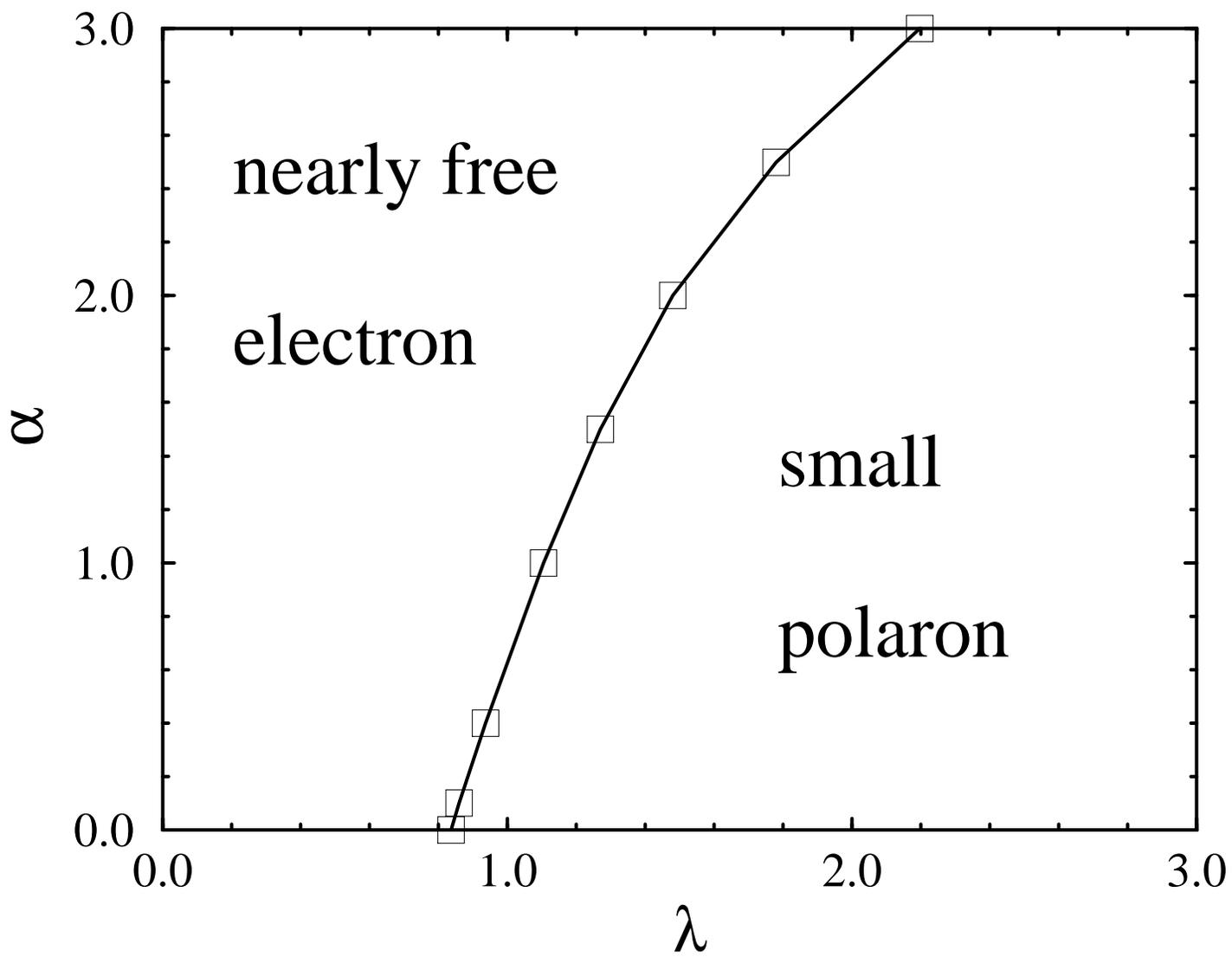
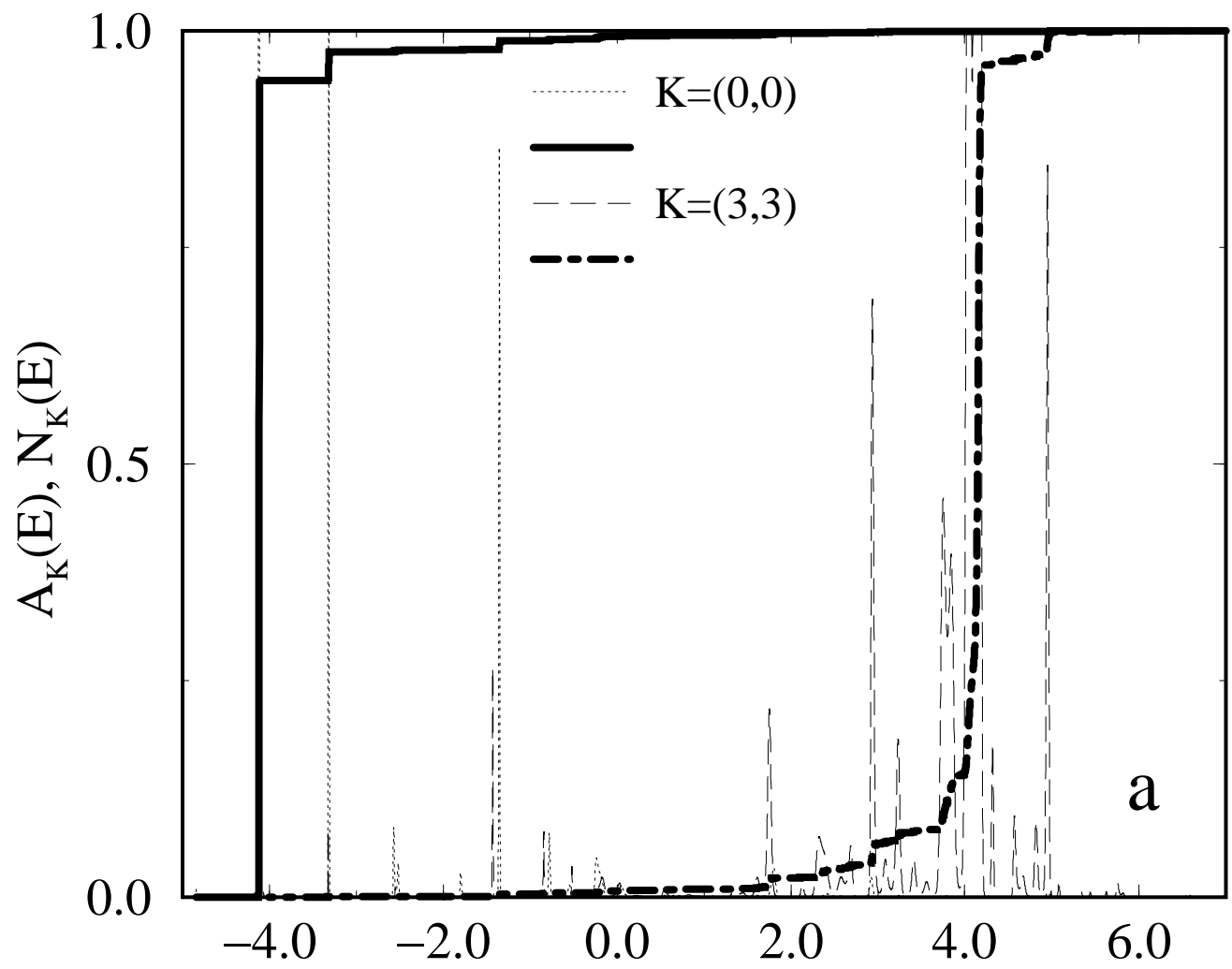
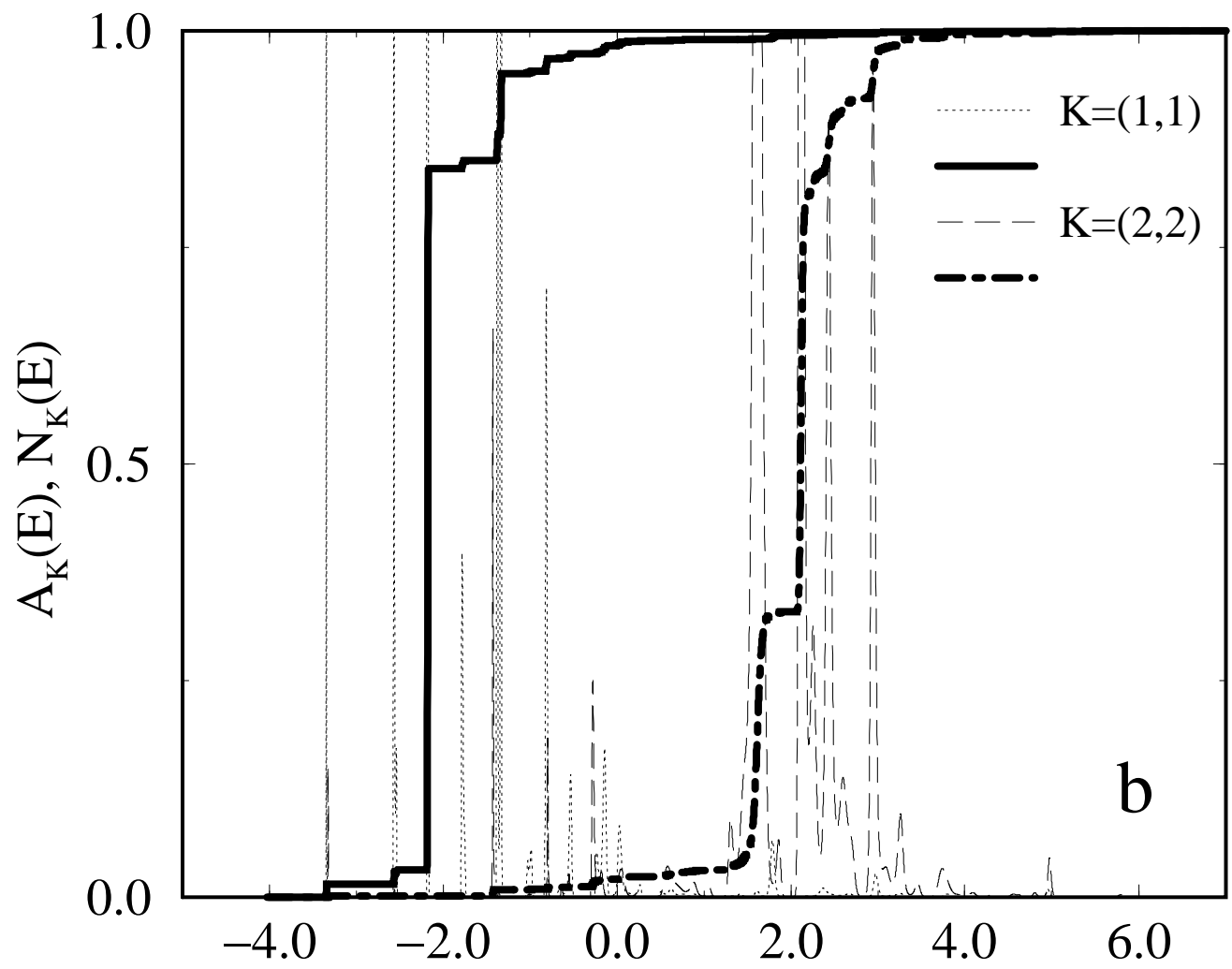
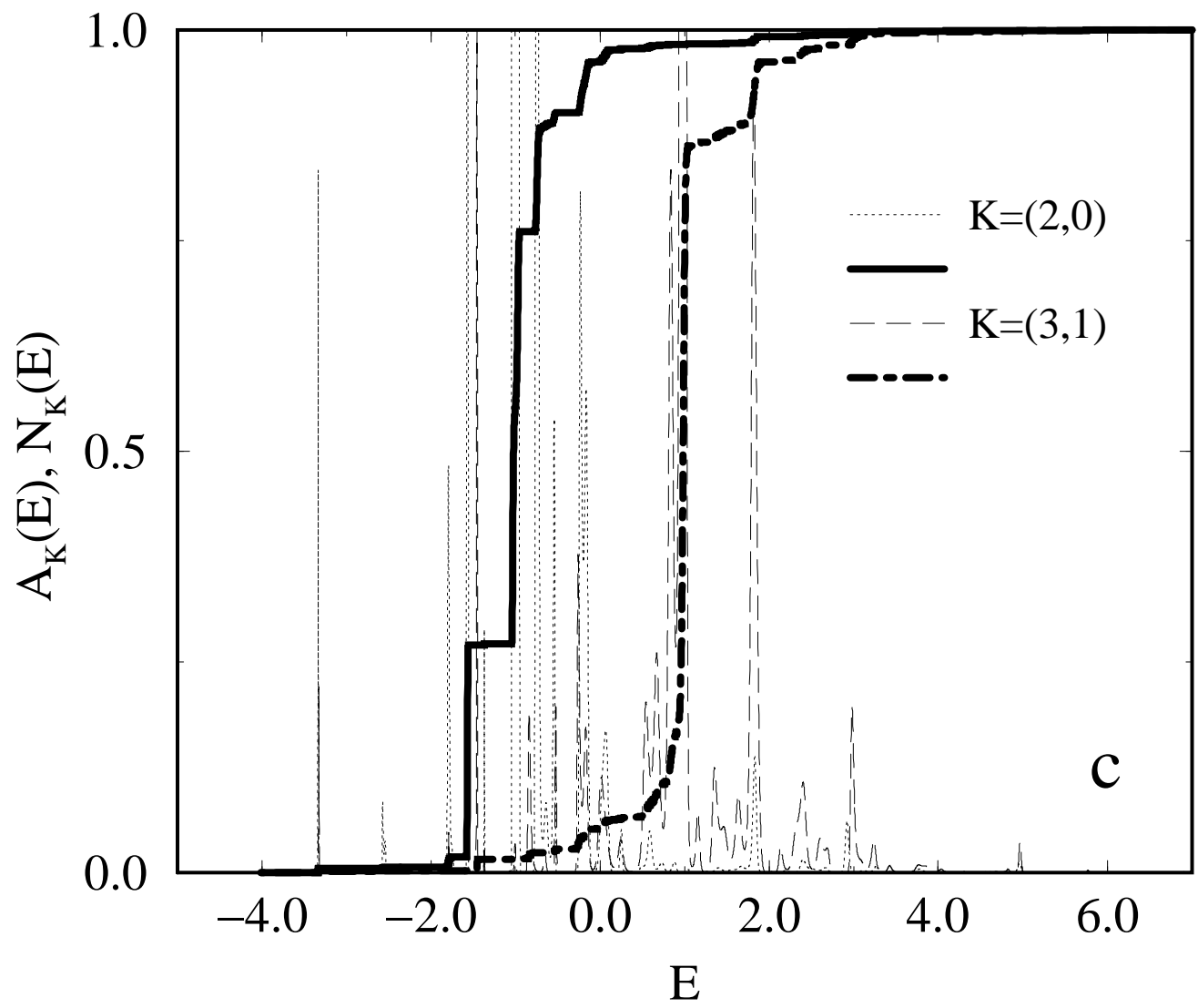
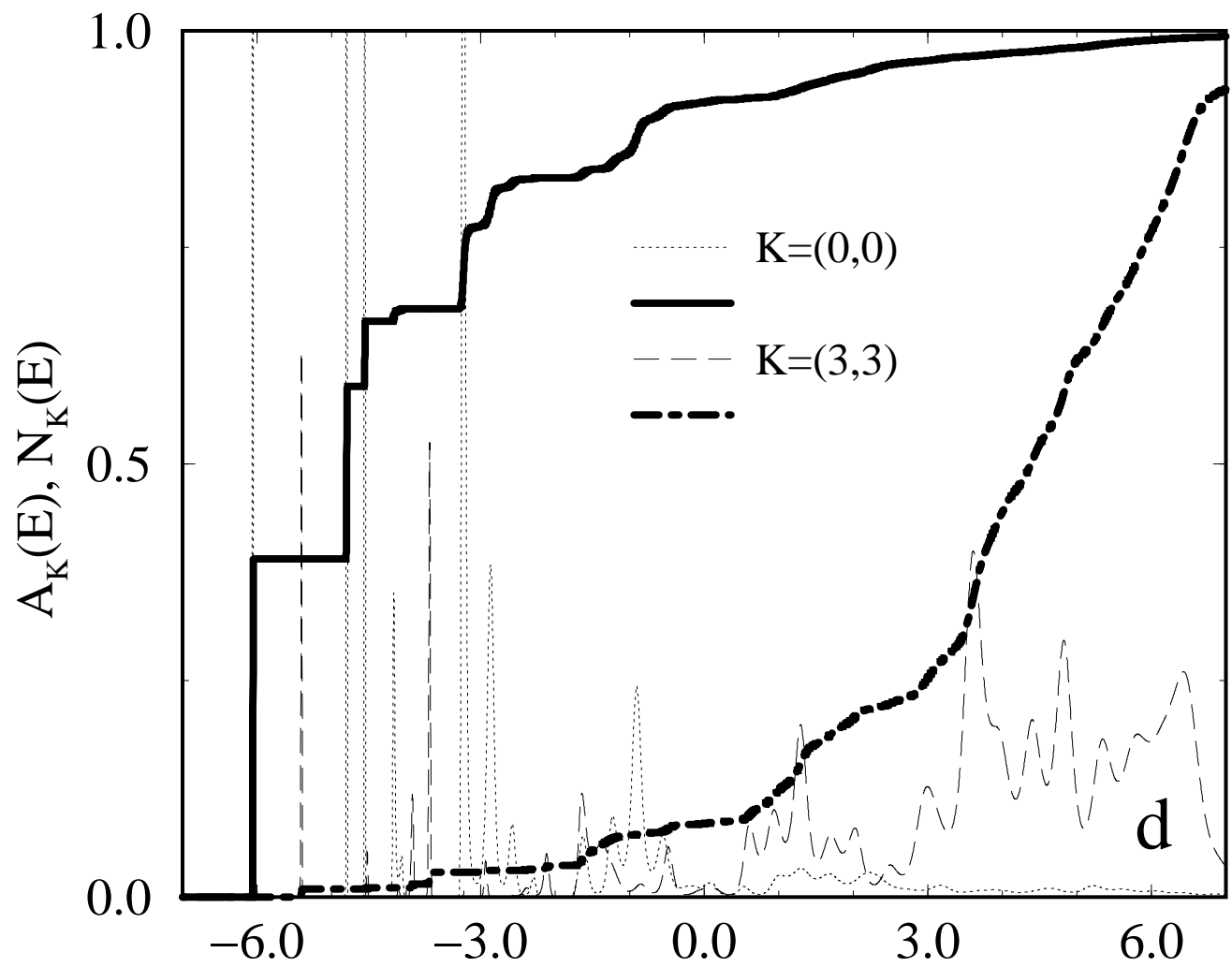
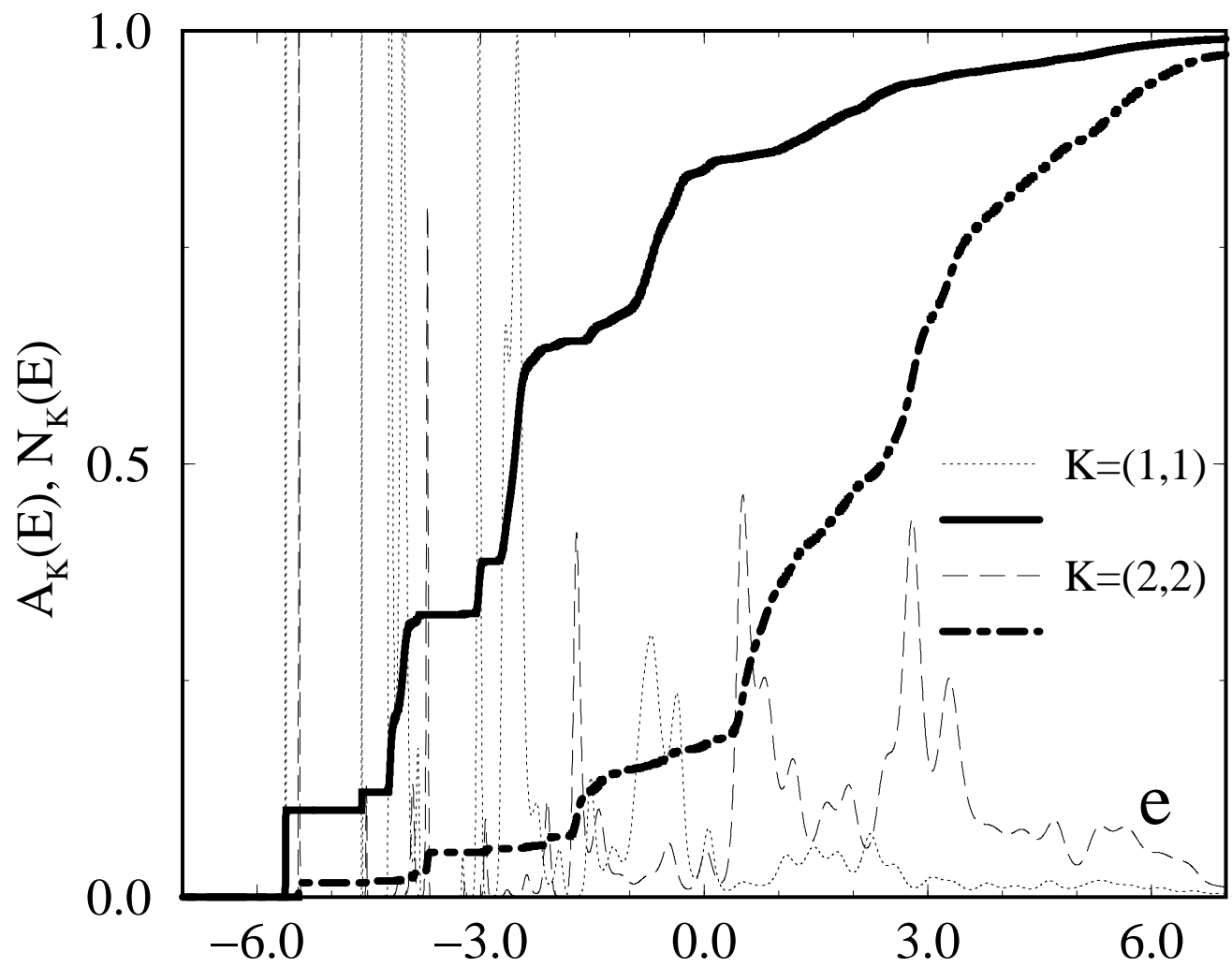
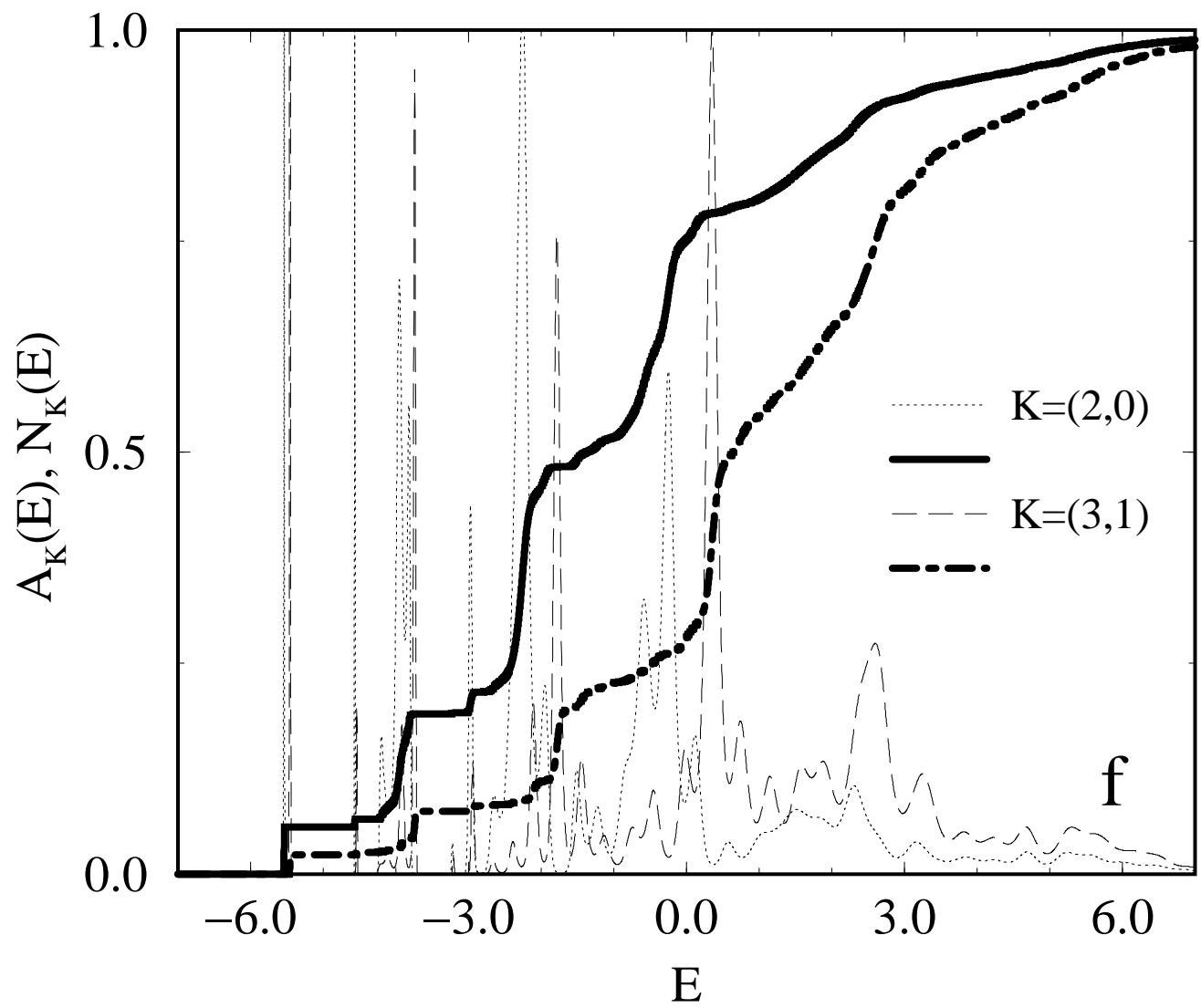
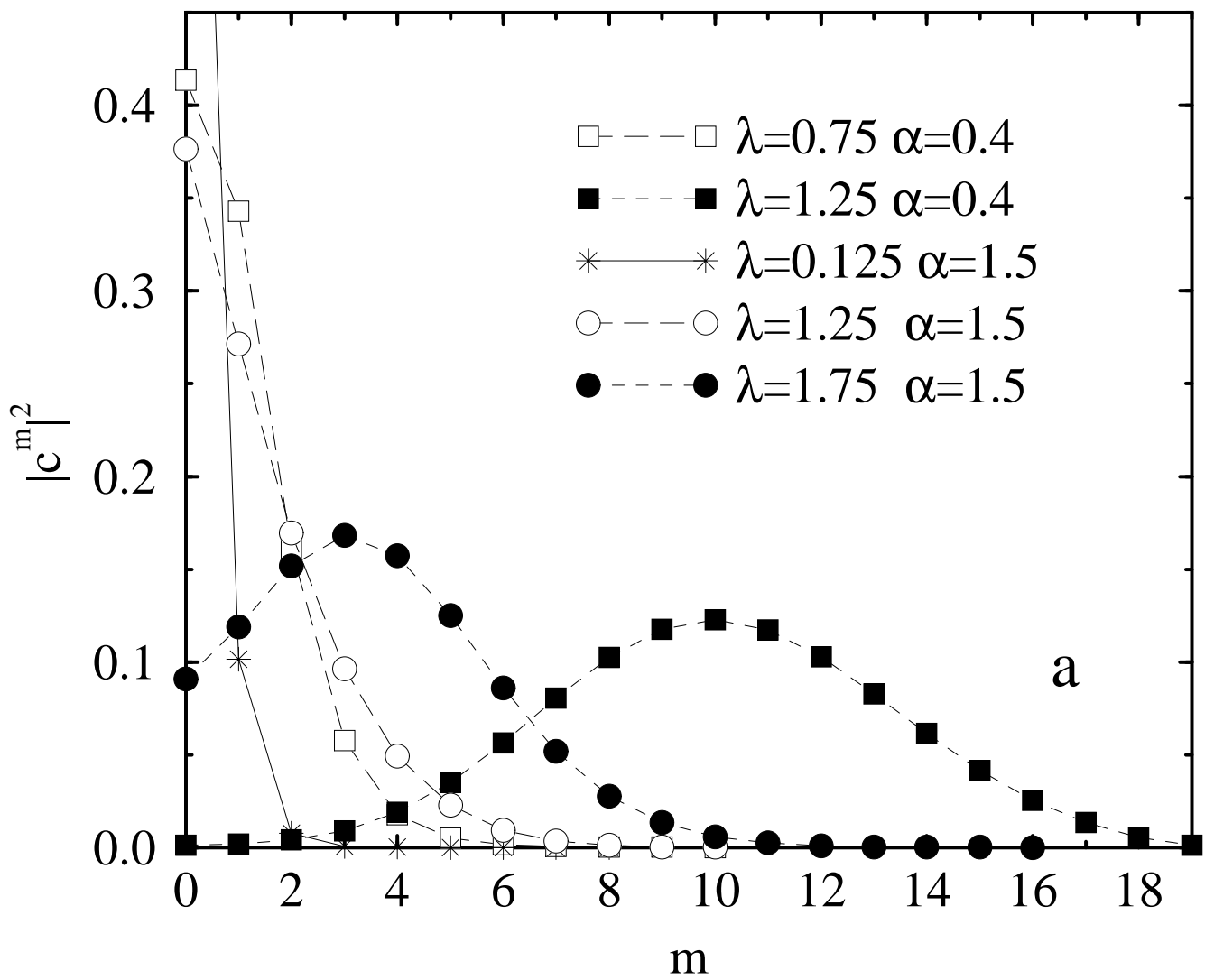
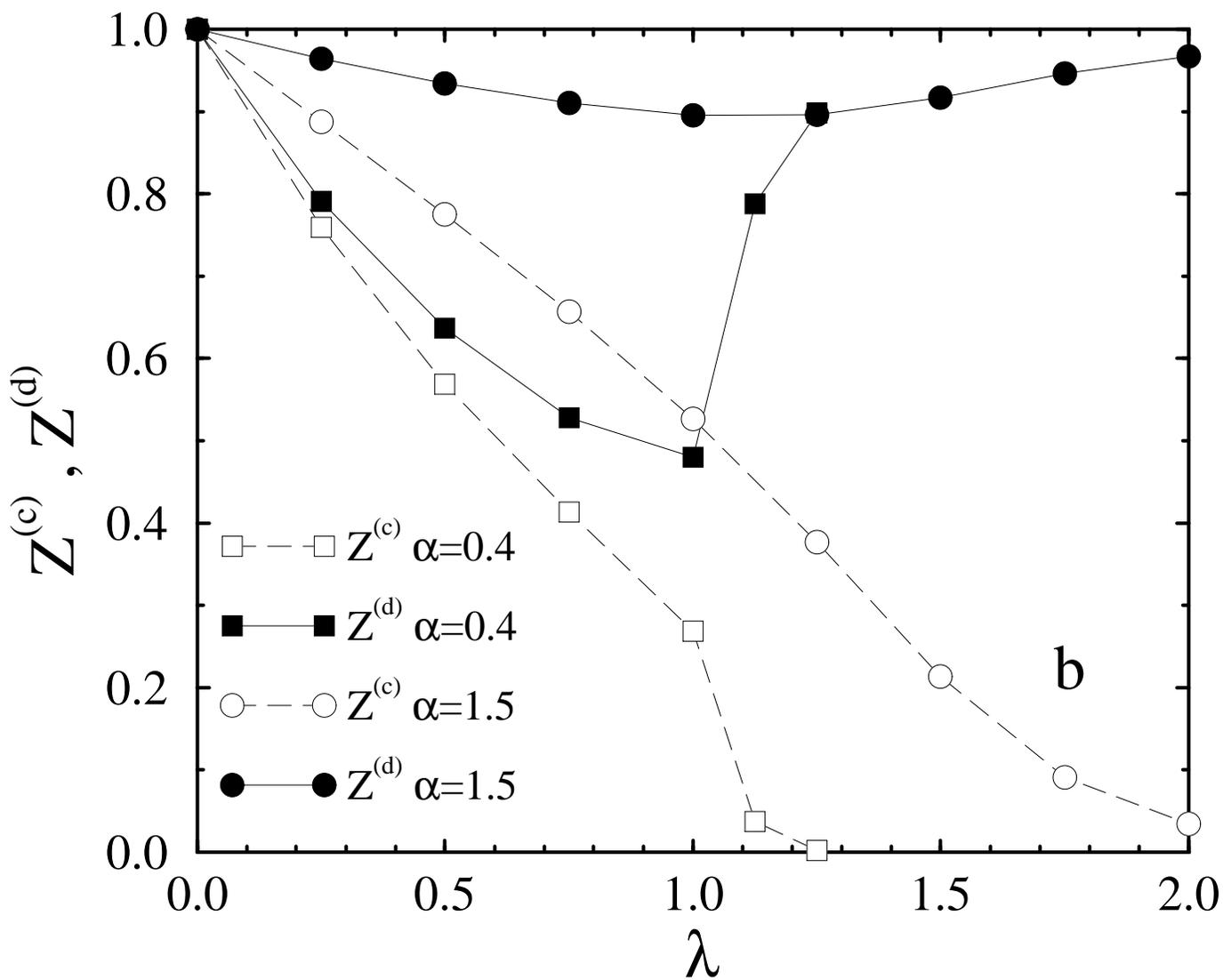
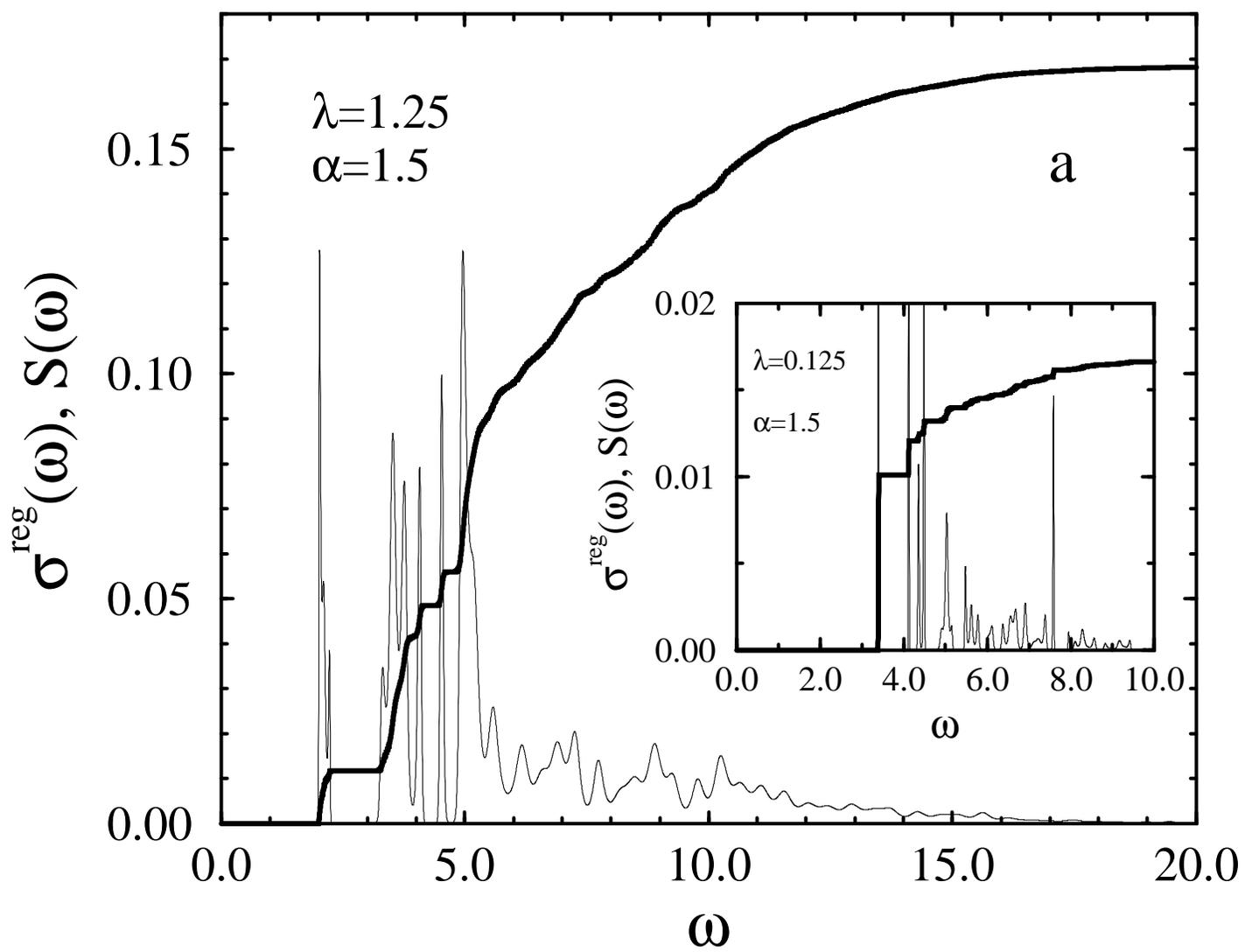
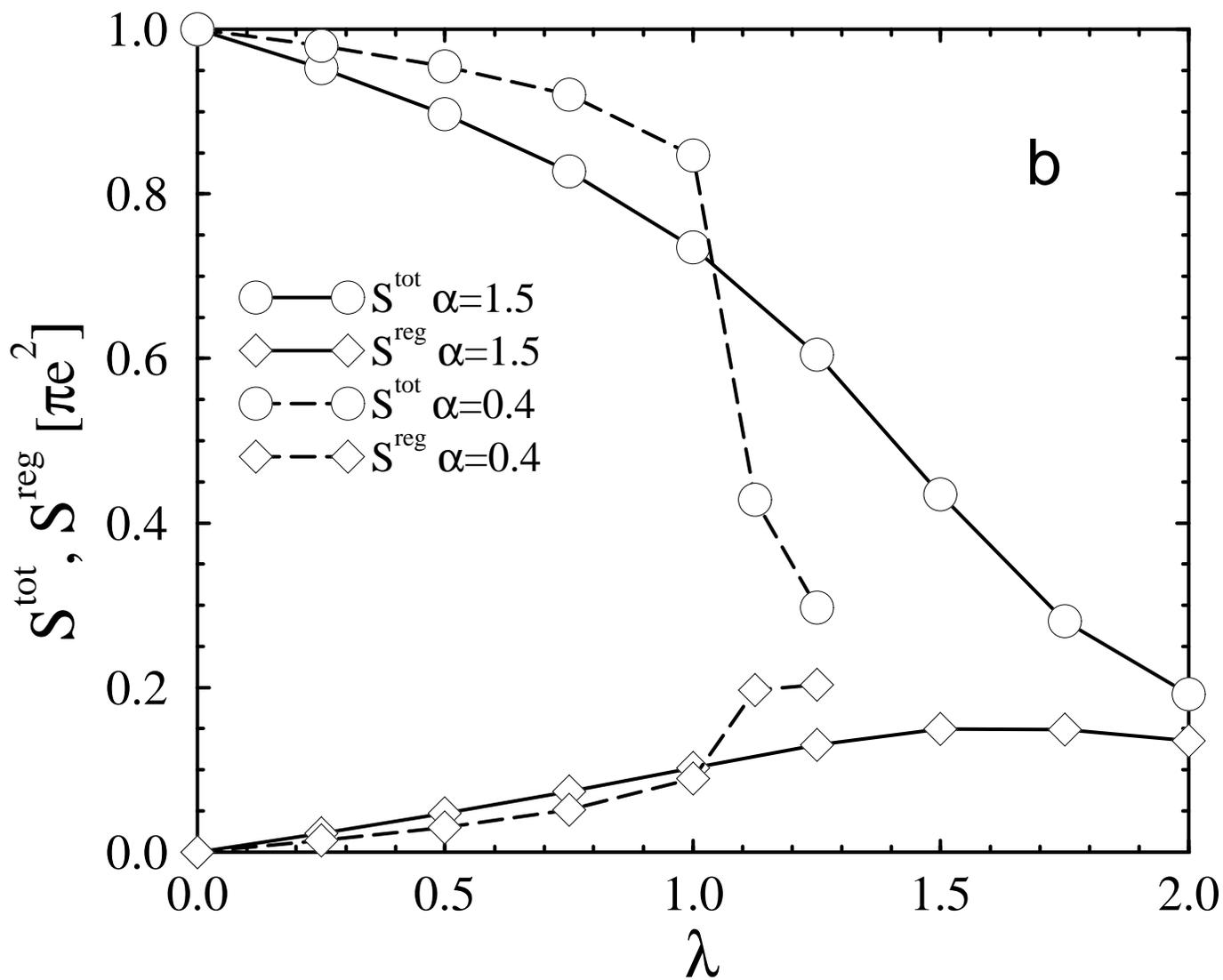
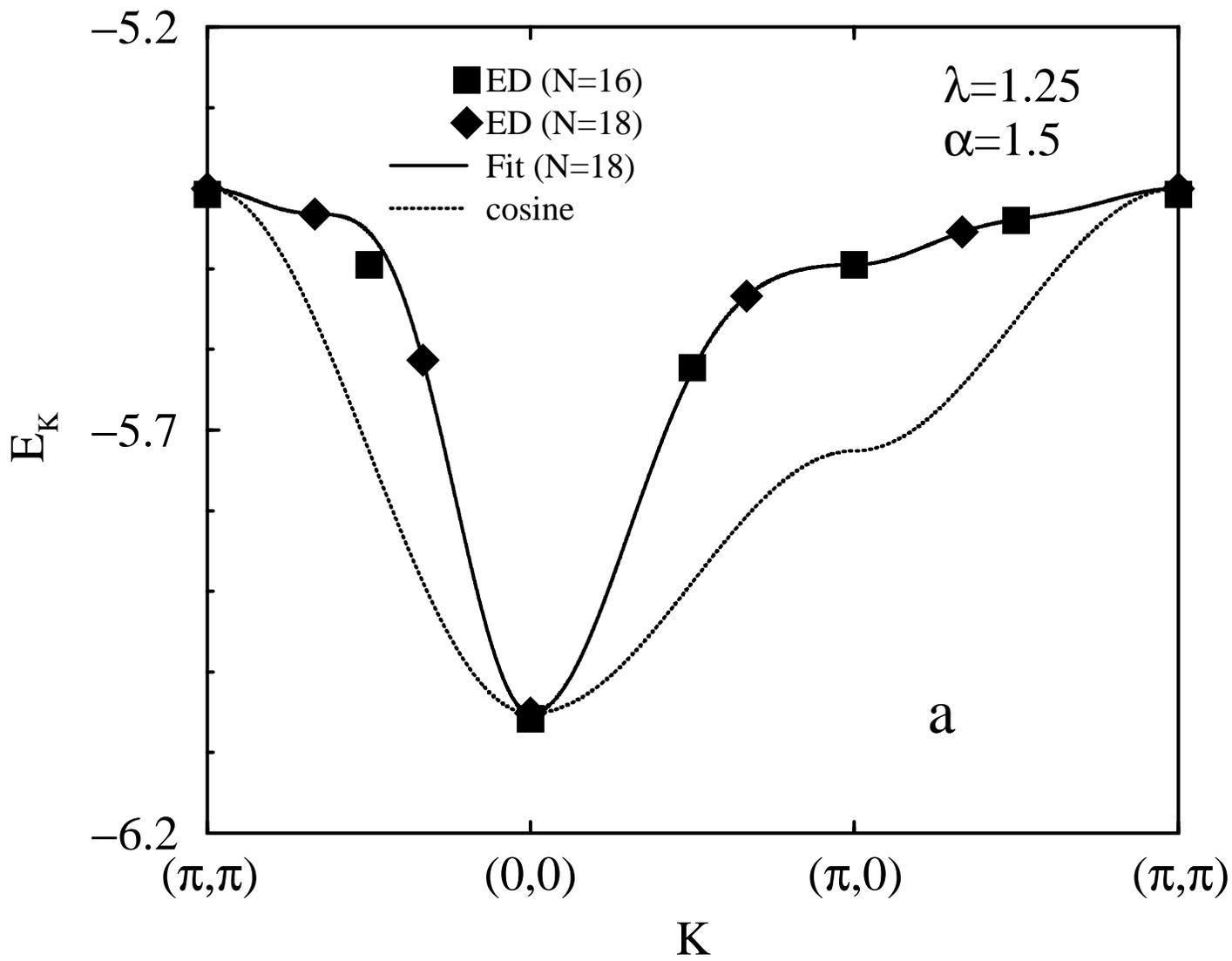
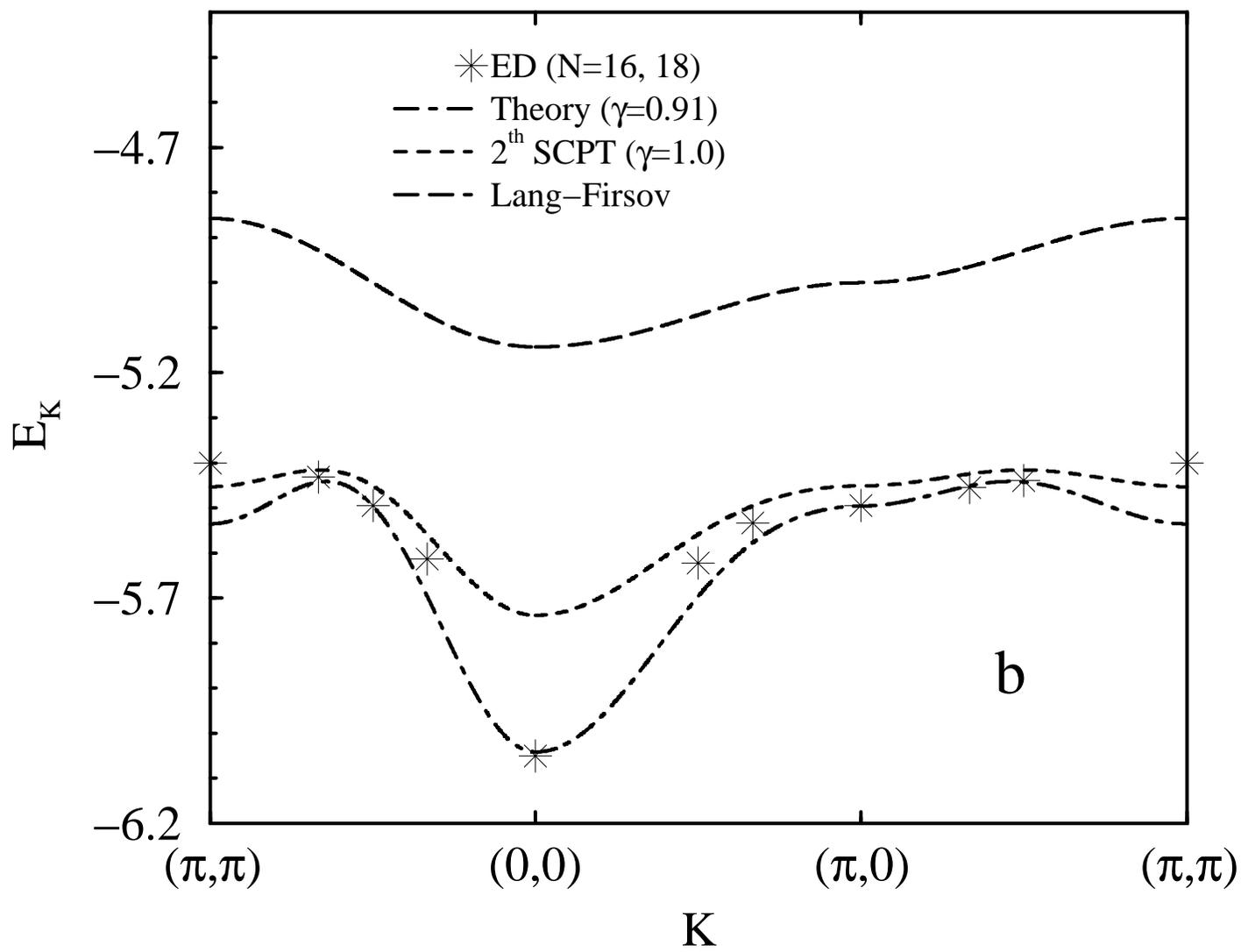
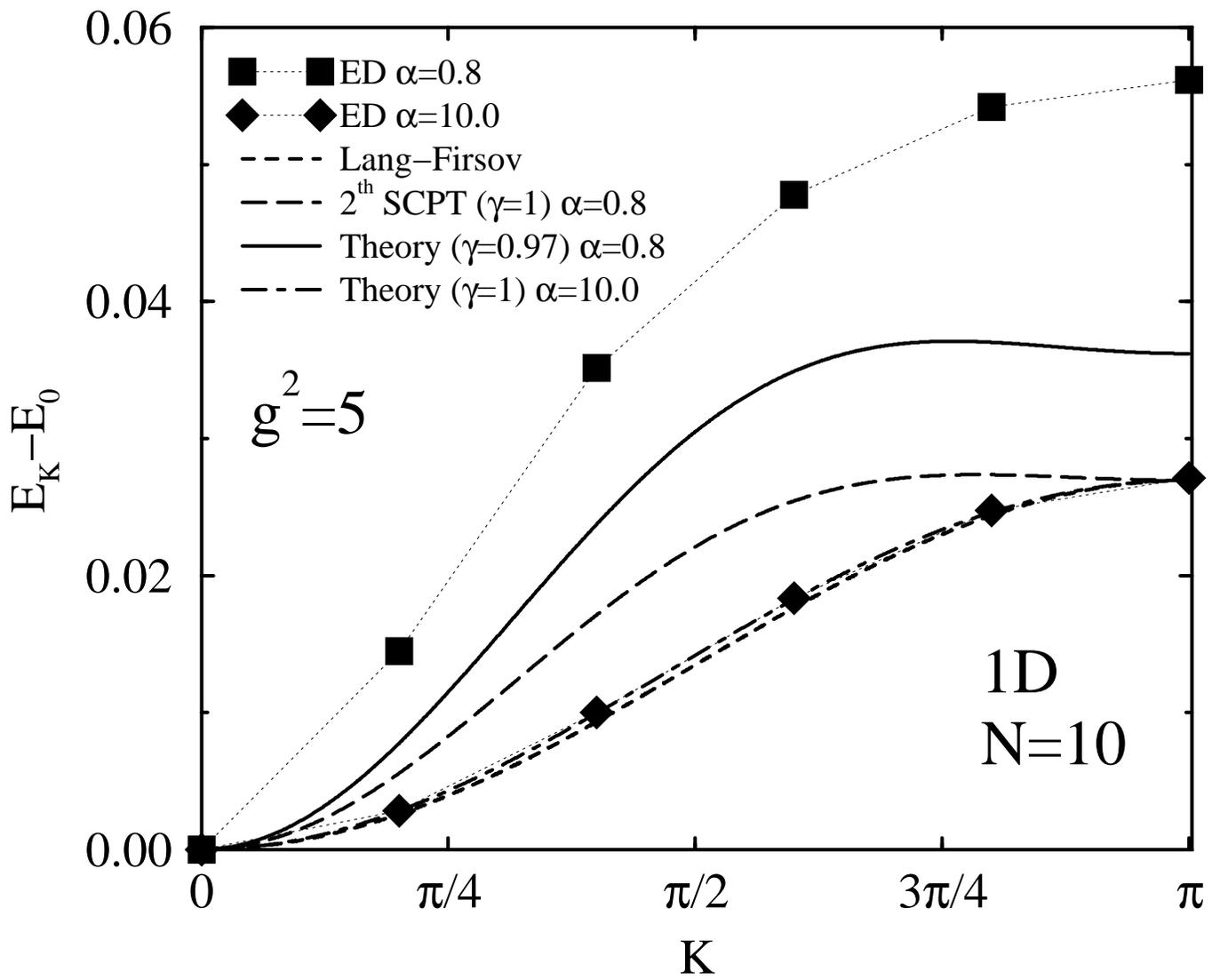
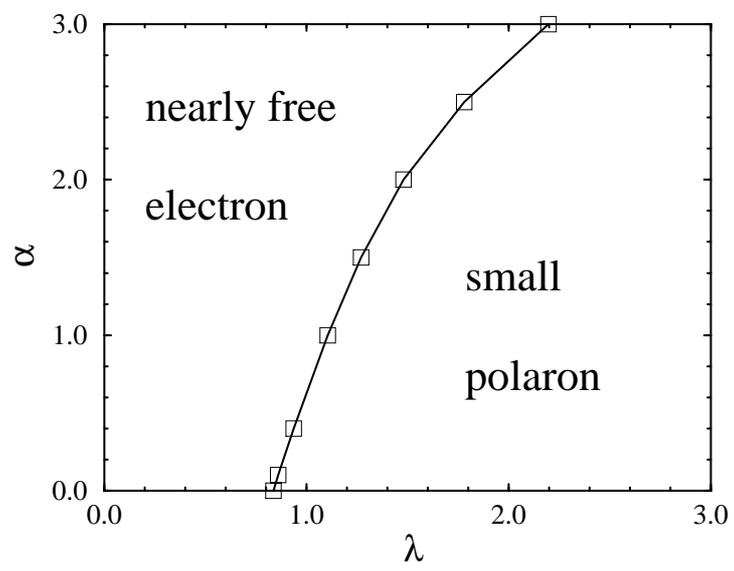
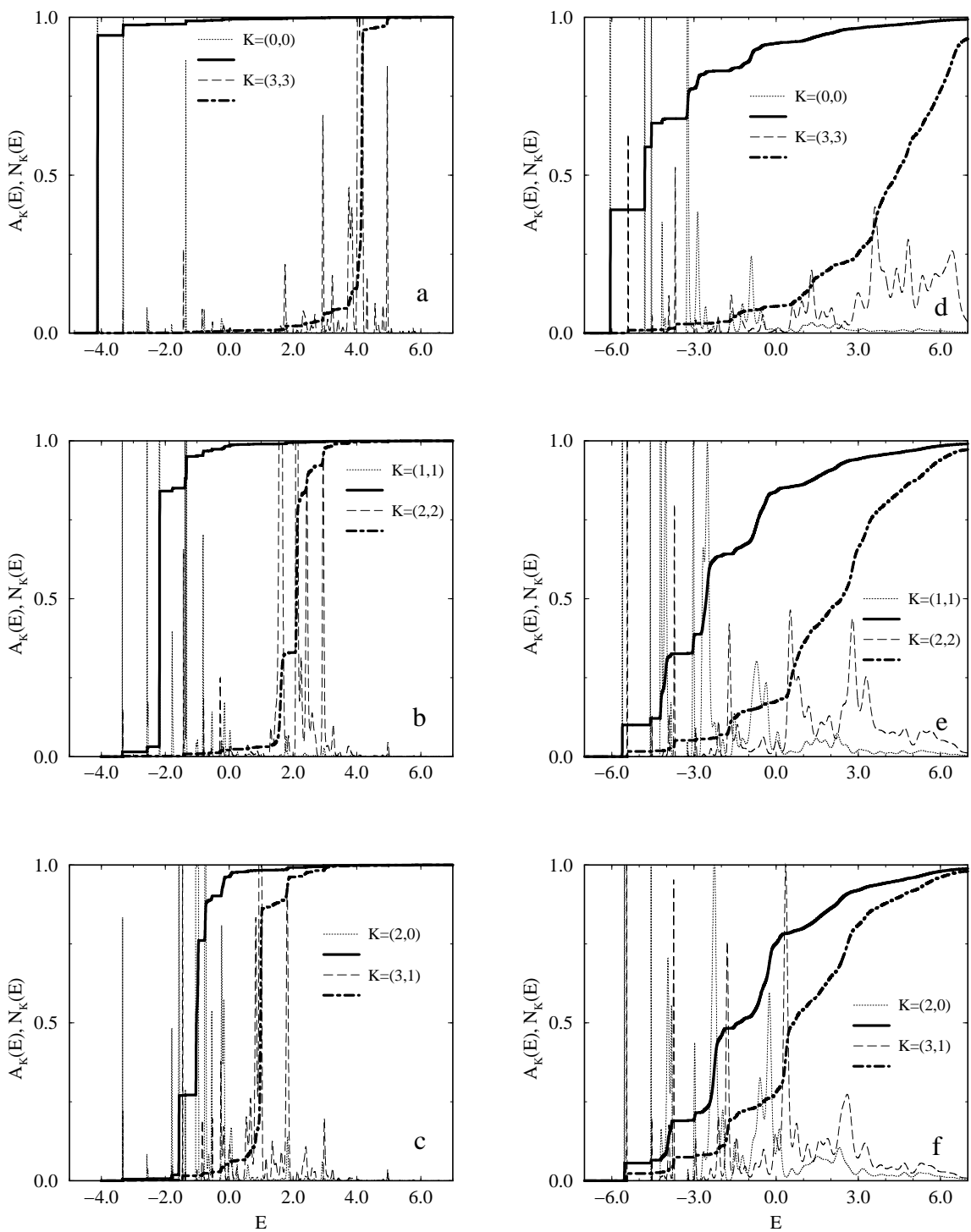
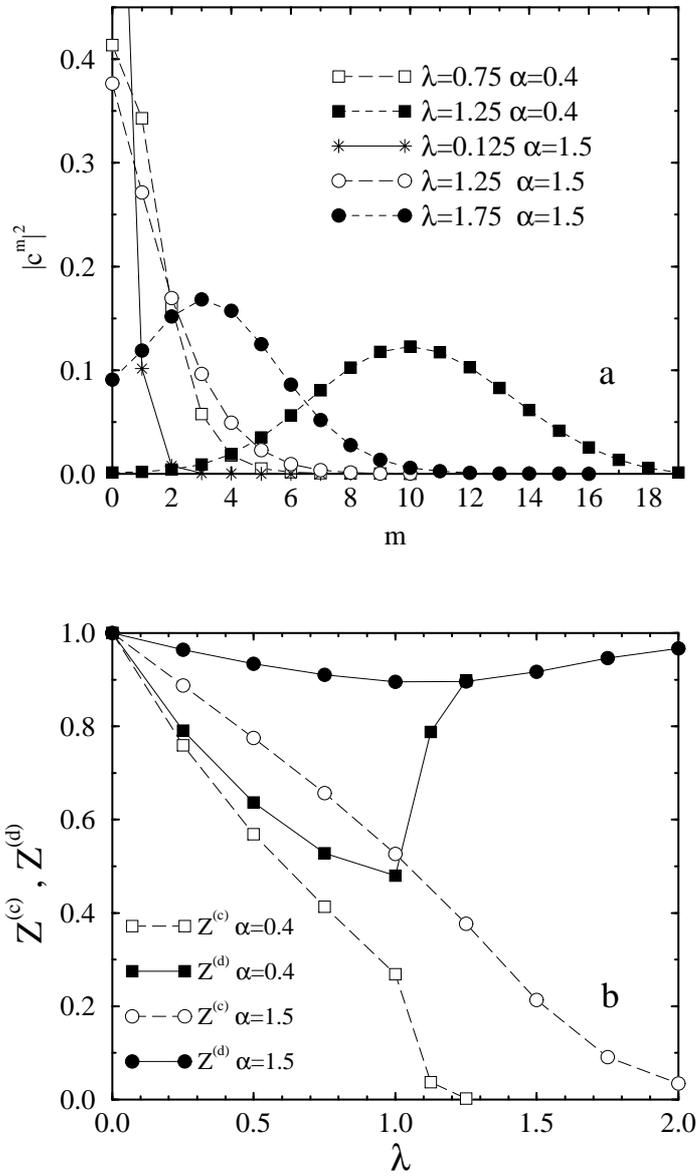
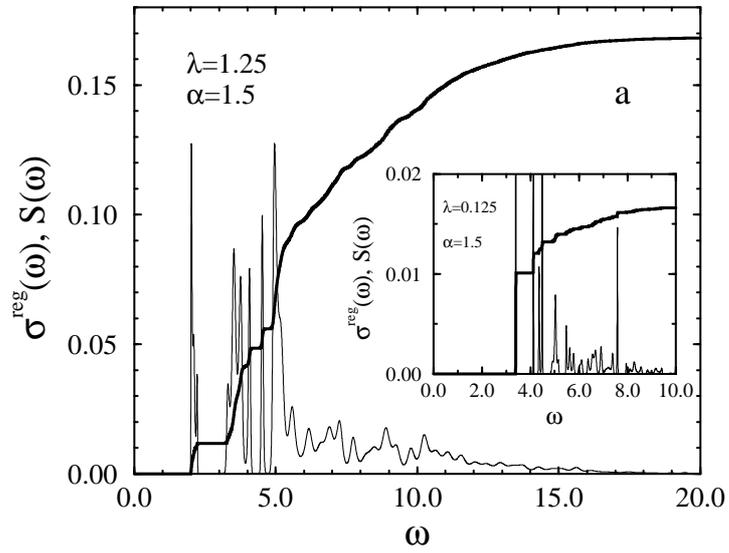
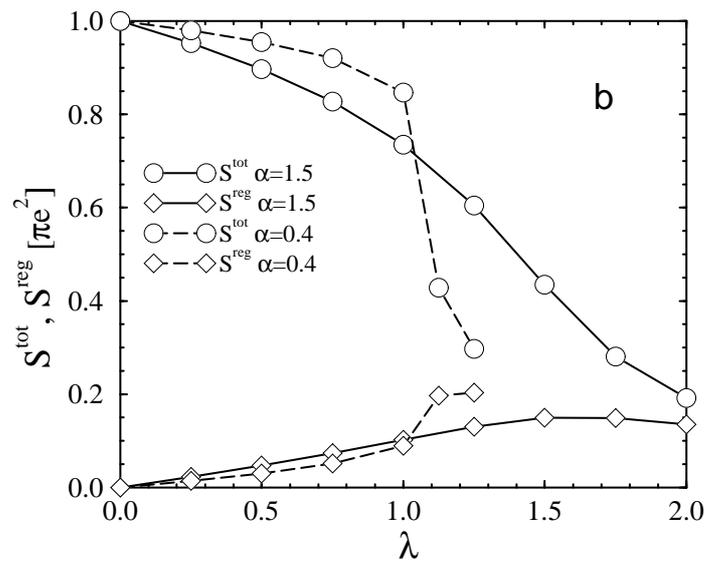
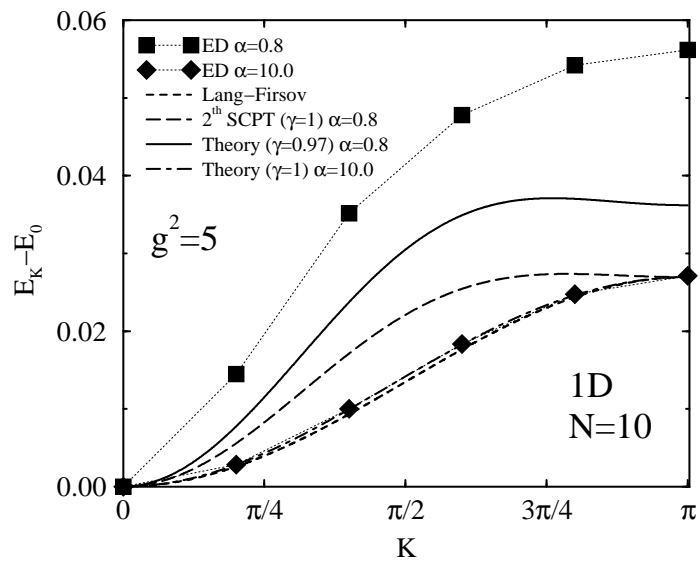
\end{document}